\documentclass[collapsecite]{iopart} % PRINTED FORMAT

\usepackage{graphicx}

\begin{document}

\title[Inelastic Collisions of Solitary Waves in Anisotropic BECs]{%
Inelastic Collisions of Solitary Waves in Anisotropic
Bose-Einstein Condensates:
Sling-Shot Events and Expanding Collision Bubbles}
\author{C.~Becker, K. Sengstock}
\address{%
Institut f\"ur Laser-Physik, Universit\"at Hamburg,
Luruper Chaussee 149, 22761 Hamburg, Germany}
\author{P.\ Schmelcher}
\address{%
Zentrum f\"ur Optische Quantentechnologien, Universit\"at
Hamburg, Luruper Chaussee 149, 22761 Hamburg, Germany}
\author{P.G.~Kevrekidis}
\address{%
Department of Mathematics and Statistics, University of Massachusetts,
Amherst, Massachusetts 01003-4515, USA}
\author{R.\ Carretero-Gonz\'alez\footnote{%
Author to whom correspondence should be addressed.}
}
\address{%
Nonlinear Dynamical Systems Group,%
\footnote{\texttt{URL:} http://nlds.sdsu.edu/}
Computational Science Research Center, and Department of Mathematics and
Statistics, San Diego State University, San Diego, CA 92182-7720, USA}
%
%\ead{cbecker@physnet.uni-hamburg.de}
%\ead{rcarretero@mail.sdsu.edu}


\begin{abstract}

We study experimentally and theoretically
the dynamics of apparent dark soliton stripes in an
elongated Bose-Einstein condensate.
% referring to a recent experimental setup
%for a single repulsive component (C. Becker {\it et al.},
%Nature Phys. {\bf 4}, 496 (2008)) \cite{Becker:Nature:2008}.
We show that for the trapping strengths corresponding
to our experimental setup, the transverse confinement along one
of the tight directions is not strong enough to arrest the formation
of solitonic vortices or vortex rings.
These solitonic vortices and vortex rings, when integrated along the transverse
direction, appear as dark soliton stripes along the longitudinal direction thereby
hiding their true character. The latter significantly
modifies the interaction dynamics during collision events and
can lead to apparent examples of inelasticity and what may appear
experimentally even as a merger of two dark soliton stripes. We explain this feature by means of the interaction
of two solitonic vortices leading to a sling shot event with one of the solitonic vortices being
ejected at a relatively large speed. Furthermore we observe
expanding collision bubbles which consist of repeated inelastic
collisions of a dark soliton stripe pair with an {\it increasing} time interval
between collisions.

%
%{\bf
%All the movies corresponding to the different
%figures included in this draft can be found at:
%}
%\smallskip
%
%{\tt http://nonlinear.sdsu.edu/$\sim$carreter/noticeboard/DS\underline{~}chasing/}
%
\end{abstract}

\maketitle

%\tableofcontents

%%%%%%%%%%%%%%%%%%%%%%%%%%%%%%%%%%%%%%%%%%%%%%%%%%%%%%%%
\section{Introduction.}
%%%%%%%%%%%%%%%%%%%%%%%%%%%%%%%%%%%%%%%%%%%%%%%%%%%%%%%%

Over the past fifteen years, the advent of Bose-Einstein condensates (BECs)
has created unprecedented possibilities for the examination of phenomena
involving nonlinear waves~\cite{book1,book2}.
Ranging from bright solitary waves~\cite{expb1,expb2,expb3} and gap matter
waves~\cite{gap} to excitations of defocusing (repulsively interacting)
media most notably dark solitons~\cite{emergent,djf},
vortices~\cite{emergent,fetter1,fetter2}, as well as solitonic
vortices and vortex rings~\cite{komineas_rev},
the exploration of these phenomena has attracted considerable
theoretical as well as experimental interest.
%
%RCG: THE FOLLOWING WAS REPLACED BY THE ABOVE PHRASE:
%Although, bright solitary
%waves~\cite{expb1,expb2,expb3} and even gap matter waves~\cite{gap}
%have had their share of experimental and of course also wide
%theoretical interest, it has been chiefly the homogeneous
%excitations of defocusing (self-repulsive) media that have, arguably,
%attracted the most attention, most notably dark solitons~\cite{emergent,djf}
%and vortices~\cite{emergent,fetter1,fetter2}, as well as
%solitonic vortices and vortex rings~\cite{komineas_rev}.

Early experiments on dark solitons~\cite{han1,nist,dutton,han2} were,
at least in part, limited by the lifetimes
of these states under the influence of dynamical
instabilities in higher dimensional settings or the
effect of thermal fluctuations at temperatures closer
to the transition temperature. Nevertheless,
more recent experiments have been able to produce
a significantly increased experimental
control~\cite{engels,Becker:Nature:2008,hambcol,kip,andreas,jeffs}.
The resulting combination of sufficiently low temperatures with
---in some of the cases--- more quasi-one-dimensional regimes
has led to clear-cut observations of oscillating and interacting
dark solitons, bearing good agreement with theoretical predictions
for these structures.

Vortices, in turn, constitute the quasi-two-dimensional
generalization of dark solitons.
Following a ground breaking experiment where vortices were
produced using a phase-imprinting method~\cite{Matthews99},
other experiments produced these structures by
using a phase-imprinting method between two hyperfine
spin states of a $^{87}$Rb BEC \cite{Williams99}, they were
produced by stirring of the BECs \cite{Madison00} above a
certain critical angular speed \cite{Recati01,Sinha01,corro,Madison01}.
%which, in turn,
This led to the production of few vortices
\cite{Madison01} and even of very robust vortex lattices \cite{Raman}.
Such states were also produced by other methods including
dragging obstacles through the BEC \cite{kett99,bpaprl2}, the nonlinear interference
of condensate fragments \cite{BPAPRL}, or even
the use of the Kibble-Zurek
mechanism~\cite{bpanat,dsh_science}. Vortices of higher charge were
also produced and their dynamical instability was considered~\cite{S2Ket}.

The three-dimensional (3D) generalization of the above states consists
of multiple possibilities. The most prominent among them are solitonic
vortices (or vortex lines)
and vortex rings~\cite{emergent,komineas_rev},
both of which emerge from the instability of the dark soliton
stripe ---the multi-dimensional analog of a 1D dark
soliton~\cite{komineas_pra}. A solitonic vortex is the 3D extension of a two-dimensional
(2D) vortex by (infinitely and homogeneously)
extending the solution into the axis perpendicular to the vortex plane.
In a realistic experimental setup, which requires an external trapping
potential, the solitonic vortices naturally acquire a finite length.
In that case, solitonic vortices are vorticity ``tubes'' that are straight across
the BEC cloud or bent in U or S shapes depending on the aspect ratio of
the BEC cloud \cite{VR:USshapedVLs1,VR:USshapedVLs2}.
If a solitonic vortex is bent enough to close on itself or if two
solitonic vortices are close enough to each other they can produce
a vortex ring \cite{VR:CrowInstab}.
Vortex rings are 3D structures whose core is a closed loop with vorticity
around it \cite{donelly} (i.e., a vortex that is looped back into itself).
Vortex rings can also be produced by an impurity traveling faster
than the speed of sound of the background \cite{rcg:62N};
by nonlinear interference between colliding blobs
of the background material \cite{rcg:55,ourbrian1};
by phase and density engineering techniques
\cite{VR:Dutton1,jeffs};
or even by introducing ``bubbles'' of one component in the other
component in two-component BEC systems \cite{VR:2C}.

In the present work we explore in detail how in experimentally
available settings it is possible to produce an interplay between
one-dimensional and three-dimensional phenomenologies (with the two-dimensional
case, as an interesting intermediary ---see below). This interplay
is found to be responsible for {\it dramatic}
events such as seemingly perfectly ``plastic'' collisions
(which entirely defy the near-integrable nature of
solitary waves as such). We show that such events which
we have observed in our experiments are not
aberrations but rather a direct consequence of the
hidden character of the solitons, namely their conversion
to solitonic vortices during their dynamical evolution. The
strong interaction of such solitonic vortices is found to potentially
sling shot one of them towards the background and produce
such a ``plastic'' collision which appears as a merger of two dark soliton stripes
when integrated along one spatial dimension ---as it necessarily happens in
absorption imaging in current experimental settings.
Further counter-intuitive phenomena occurring
in this setup include a sequence of apparent dark soliton stripe collisions
with increasing collision times, i.e.~a series of ``expanding collision bubbles''.
This is shown to be produced by mutual rotation of solitonic vortices around each other.

Our presentation is structured as follows. In Sec.~\ref{SEC:exp}
we present the experimental setup and the corresponding experimental
observations displaying the apparent, unexpected, merger of two dark soliton stripes.
In Sec.~\ref{SEC:theo} we introduce the theoretical model
and its corresponding numerical experiments that allow to
elucidate the nature of this unexpected merger.
We also include in this section some other intriguing cases
where solitonic vortices rotating around each other tend to slow down
their rotation frequency.
Finally, in Sec.~\ref{SEC:conc} we present our conclusions
and identify some future areas for possible exploration.

%%%%%%%%%%%%%%%%%%%%%%%%%%%%%%%%%%%%%%%%%%%%%%%%%%%%%%%%
\section{Experimental Setup and Observations}
\label{SEC:exp}
%%%%%%%%%%%%%%%%%%%%%%%%%%%%%%%%%%%%%%%%%%%%%%%%%%%%%%%%

For the experiments presented here we start by trapping $5\!\times\!10^9$
atoms of $^{87}$Rb in a magneto-optical trap which are subsequently
compressed and cooled in an optical molasses. The atoms are further
cooled by evaporative cooling in a magnetic trap before we load them
into an optical dipole trap. After a second evaporative cooling stage
over 20\,s, an elongated and almost pure BEC of about $5\!\times\!10^4$
atoms in the $|5^2S_{1/2}, F\!=\!1, m_F\!=\!-1\rangle$ state is
produced with a chemical potential of less than 20\,nK.
The trap frequencies read $\omega_{x,y,z}=2 \pi \times (85, 133, 5.9)$\,Hz.
%$(f_z,f_y,f_x)=(85,133,5.9)$Hz.
These correspond to a ``cigar-shaped'' trap elongated
along the $z$-direction and strongly confined along the
transverse $x$ and $y$ directions.
The peak density of $n_0=5.8\!\times\!10^{13} \,\textrm{cm}^{-3}$
corresponds to a speed of sound of $c_s= 1.0 \, \textrm{mm/s}$ and
a healing length of $\xi= 0.7\,\mu$m.

%%%%%%%%%%%%%%%%%%%%%%%%%%%%%%%%
\begin{figure}[htbp]
\includegraphics[width=12.5cm]{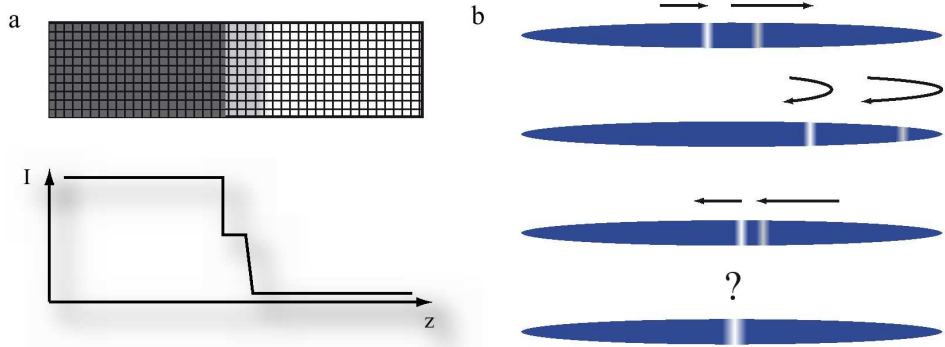}
\caption{
Experimental setup for the chase scenario for two dark
soliton stripes. Two dark soliton stripes are generated
that travel in the same direction at different velocities. After a quarter
oscillation period they reverse their direction of propagation and travel
in the opposite direction where the fast soliton will eventually
overtake its slower counterpart.
(a) Intensity profile used to generate two solitons at different
velocities. To generate a faster soliton
less phase feed is applied over a larger region.
(b) Schematic representation of the chase scenario of a fast (indicated by
a gray stripe)
and a slow (indicated by a white stripe) soliton.
}
\label{fig:exp1}
\end{figure}
%%%%%%%%%%%%%%%%%%%%%%%%

In order to seed dark soliton stripe structures in the experiment,
we employ the well-established method of optical phase imprinting
in Bose-Einstein condensates~\cite{han1,nist,Becker:Nature:2008}.
For this purpose a laser pulse of 70\,$\mu s$ duration, blue-detuned
from atomic resonance by 8\,GHz is used. We create almost arbitrary
intensity patterns employing a spatial light modulator and image
those patterns onto the BEC through a high quality objective
yielding an optical resolution of better than 2\,$\mu$m. In this
way, the number of dark soliton stripes created can be varied, as can their
individual depths, initial positions and directions of movement
be chosen over a wide range of parameters by tailoring the light
field potentials acting on the BEC accordingly.
For the experiment
described here, we image a two-step intensity profile
(see Fig.~\ref{fig:exp1}.a) onto the BEC,
thus creating one dark soliton stripe at each phase jump.
As schematically depicted in Fig.~\ref{fig:exp1}.b,
two dark soliton stripes are generated in that way which travel in the
same direction at different velocities. The shallower and thus
faster soliton starts ahead of the deeper and slower soliton.
After a quarter oscillation period the solitons reverse their
direction of propagation and travel in the opposite direction
where the fast soliton will eventually overtake its slower
counterpart. The solitons start approximately $20\,\mu$m
apart from each other, the fast advancing soliton with an
initial speed of $\dot{q}_{0}^{\mathrm{f}}=0.7 \,\bar{c}_{s}$ and the
slower soliton traveling at $\dot{q}_{0}^{\mathrm{s}}=0.62 \,\bar{c}_{s}$.
These values would correspond to depths of $n_s^{\mathrm{f}}/n_0 = 0.51$
and $n_s^{\mathrm{s}}/n_0 = 0.61$ respectively.
The associated phase slips across the nodal planes of the solitons
read $\Delta \phi^{\mathrm{f}} = 0.34 \pi$ and
$\Delta \phi^{\mathrm{s}} = 0.42 \pi $
for the fast and slow soliton respectively.

%%%%%%%%%%%%%%%%%%%%%%%%%%%%%%%%
\begin{figure}[htbp]
\includegraphics[width=13.0cm]{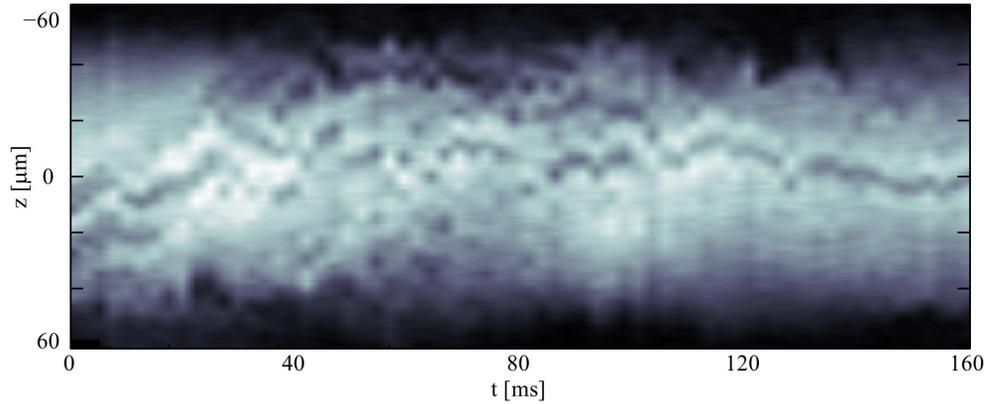}
\caption{
Experimental results depicting the chasing scenario between two dark soliton stripes.
Shown is the density plot of the condensate. Each column represents the optical
density of the elongated condensate integrated along the transverse
directions. The density depressions of the apparent dark soliton stripes
as well as the increased density of the density waves are clearly visible.
While it can not be deduced what exactly happens during the collision,
it is clearly seen that for long evolution times a single very
deep soliton that does hardly move at all is formed.
Note that we have observed structures similar
to the last 45\,ms of the graph for evolution times up to 5\,s!
}
\label{fig:exp2}
\end{figure}
%%%%%%%%%%%%%%%%%%%%%%%%

Figure~\ref{fig:exp2} shows the experimental results of the time-evolution
of two dark soliton stripes created according to the above-mentioned
experimental conditions. For each experiment the
system was let to evolve for a predetermined amount of
time and then imaged. A total of more than 60 images were
extracted 2.5\,ms apart from each other. Each image was then
integrated along the remaining transverse direction to obtain
the (integrated) 1D density along the, longitudinal, $z$ direction.
Finally, the overall evolution of the effective
1D system is rendered visible
by plotting these densities as columns in Fig.~\ref{fig:exp2}.
In the figure it is possible to see the two initial two dark soliton stripes
that travel towards the center of the trap.
Note that additional excitations of the condensate
generated during the phase imprinting process are
significantly damped prior to the soliton interaction
which ensures a better visibility of the collision process.
After the dark soliton stripes bounce back from the opposite edge of the
cloud, they strongly interact and {\em apparently merge}
after some 100\,ms.
Although the variation from one experiment to the next does
not allow for a clear depiction of the dynamics during the
collision process, we systematically obtained a resulting
cloud with a single apparent dark soliton stripes in it.
This attests to the robustness of the process.

%{\bf CB: Here we have to discuss what we claim: Comparison to
%simulations? Bending of the soliton plane during the first 0-5\,ms?
%Strong indications for the formation of vortex rings (strong
%density modulation at the edges of the condensate, much less
%in the central part) after a few ms? Formation of strongly
%modulated and very stable Vortex rings or even single
%vortices being the "final" state?
%}

%%%%%%%%%%%%%%%%%%%%%%%%%%%%%%%%
\begin{figure}[htbp]
\includegraphics[width=3.8cm]{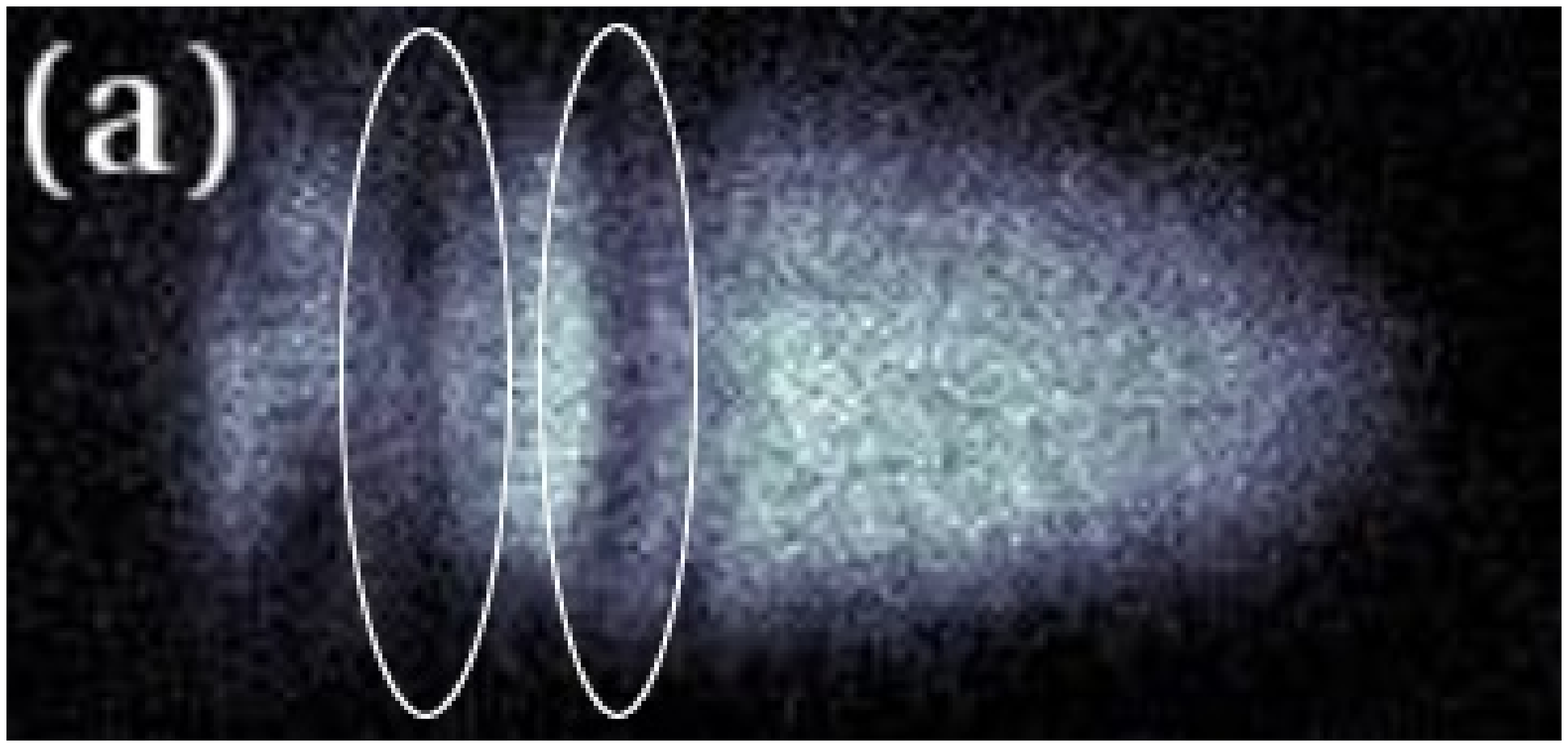}
\includegraphics[width=3.8cm]{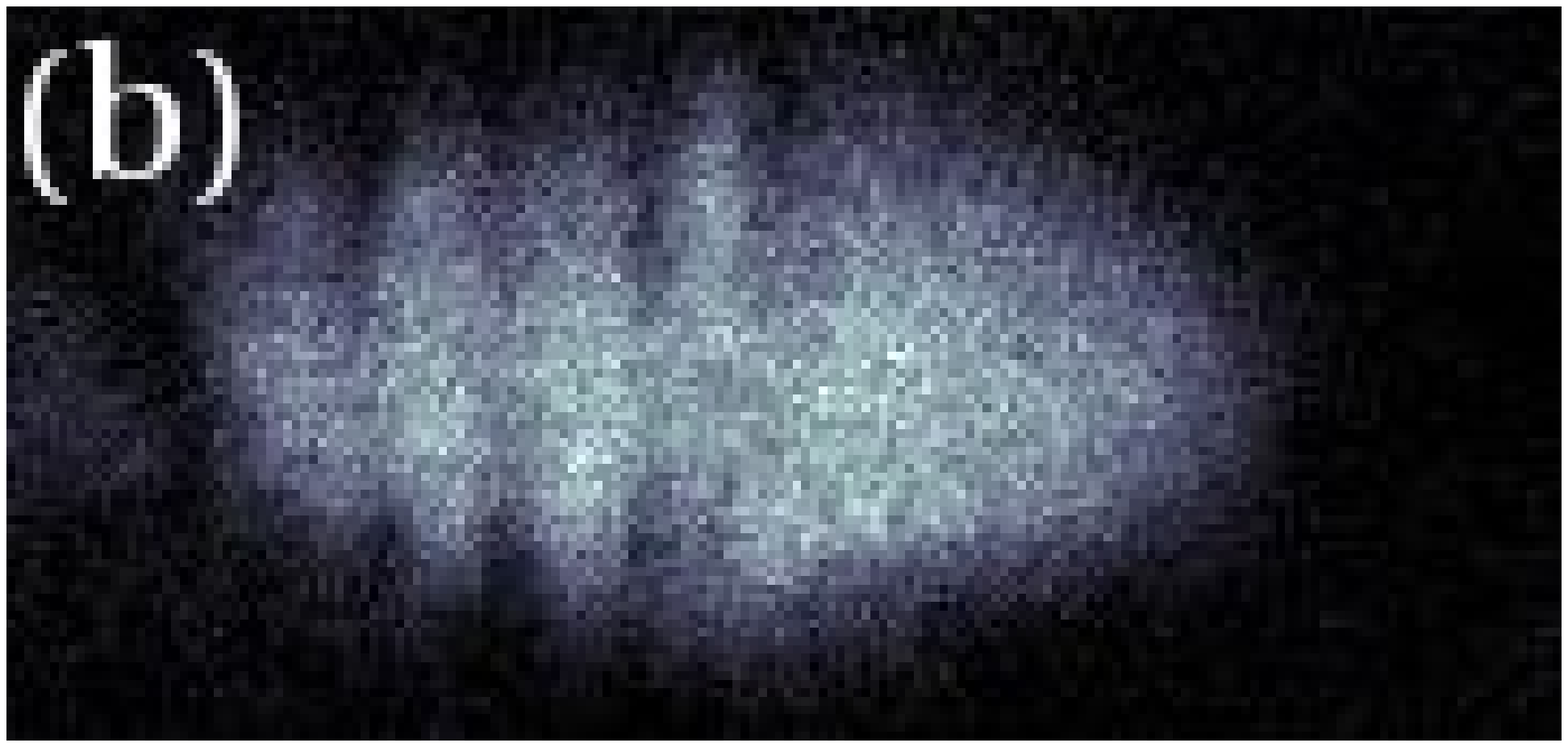}
\includegraphics[width=3.8cm]{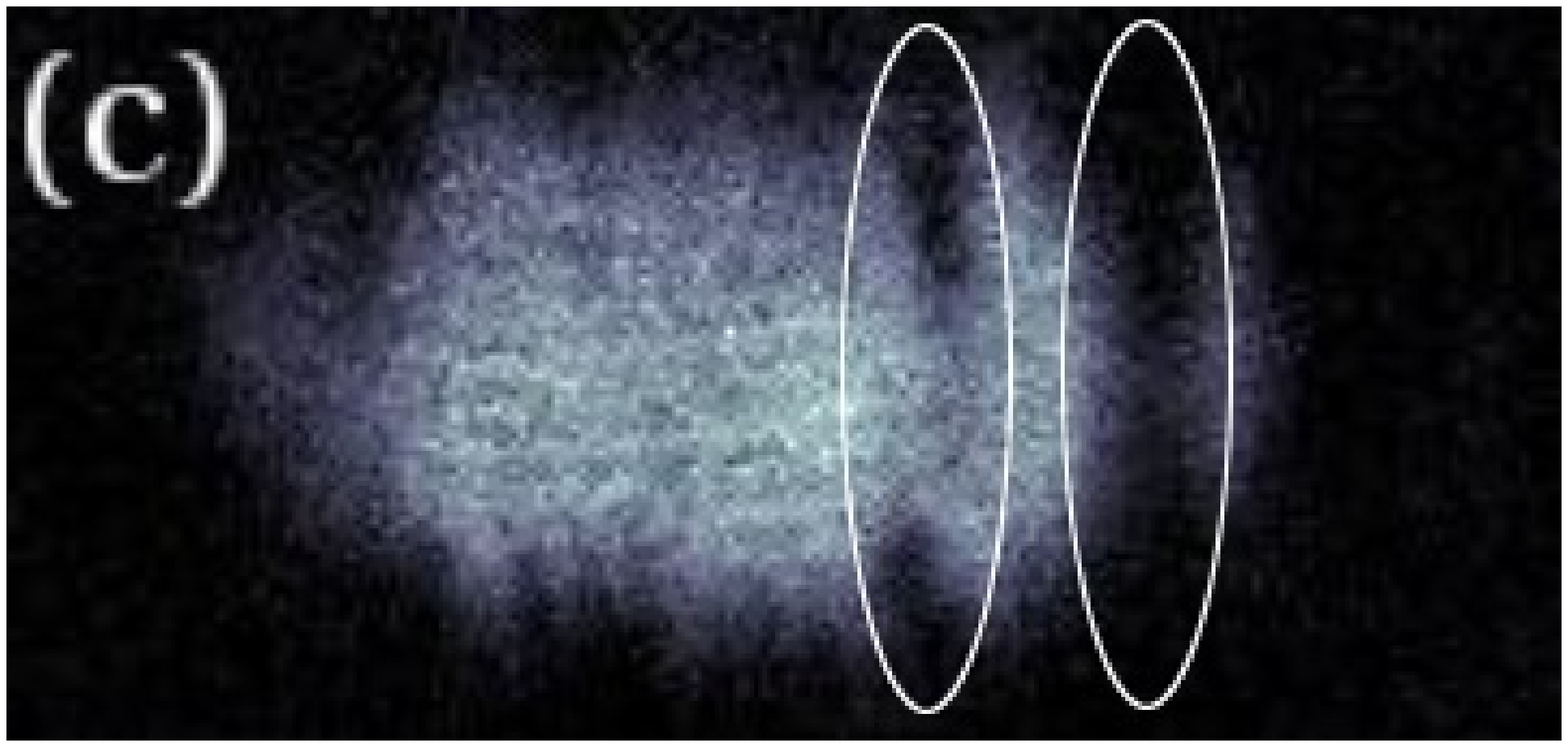}
\\
\includegraphics[width=3.8cm]{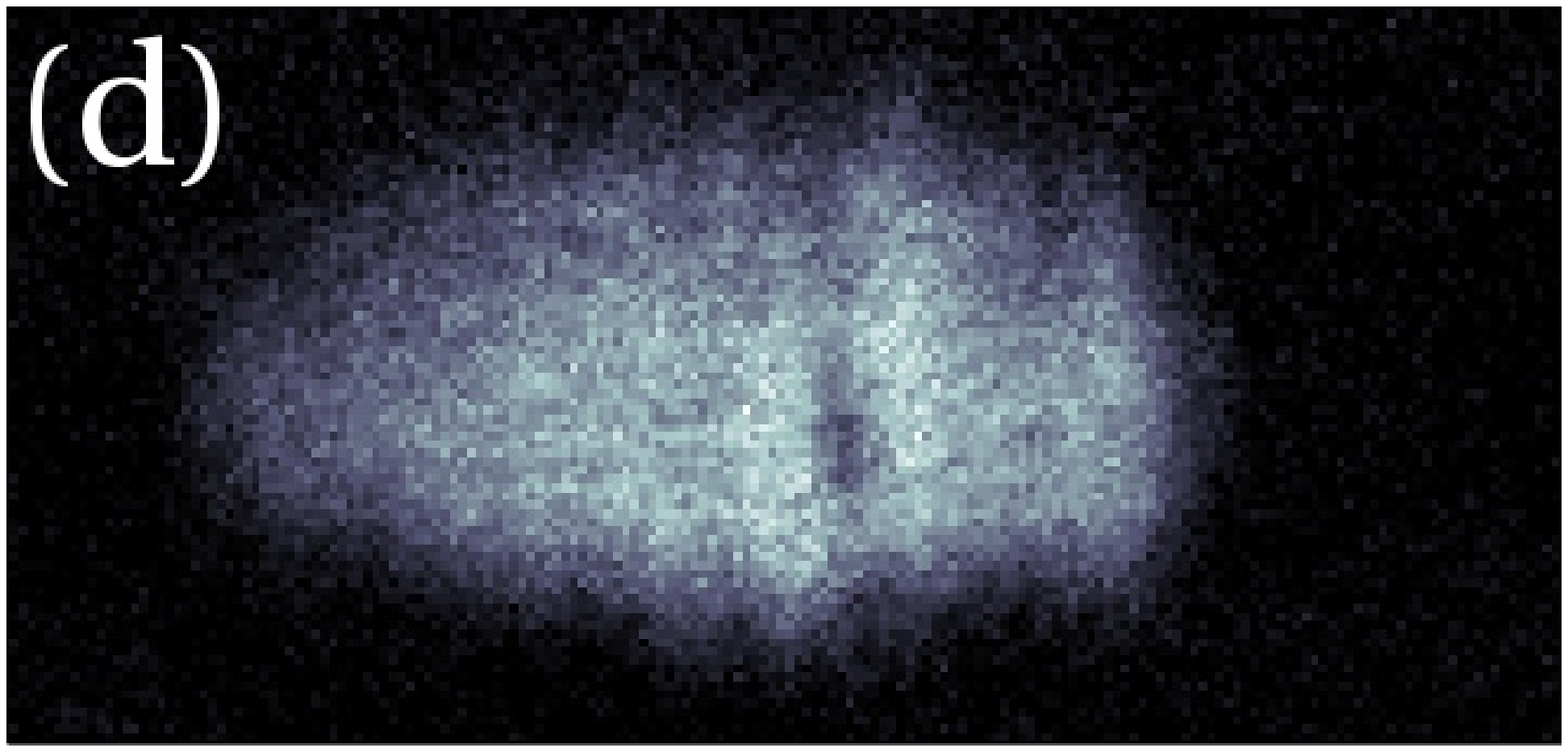}
\includegraphics[width=3.8cm]{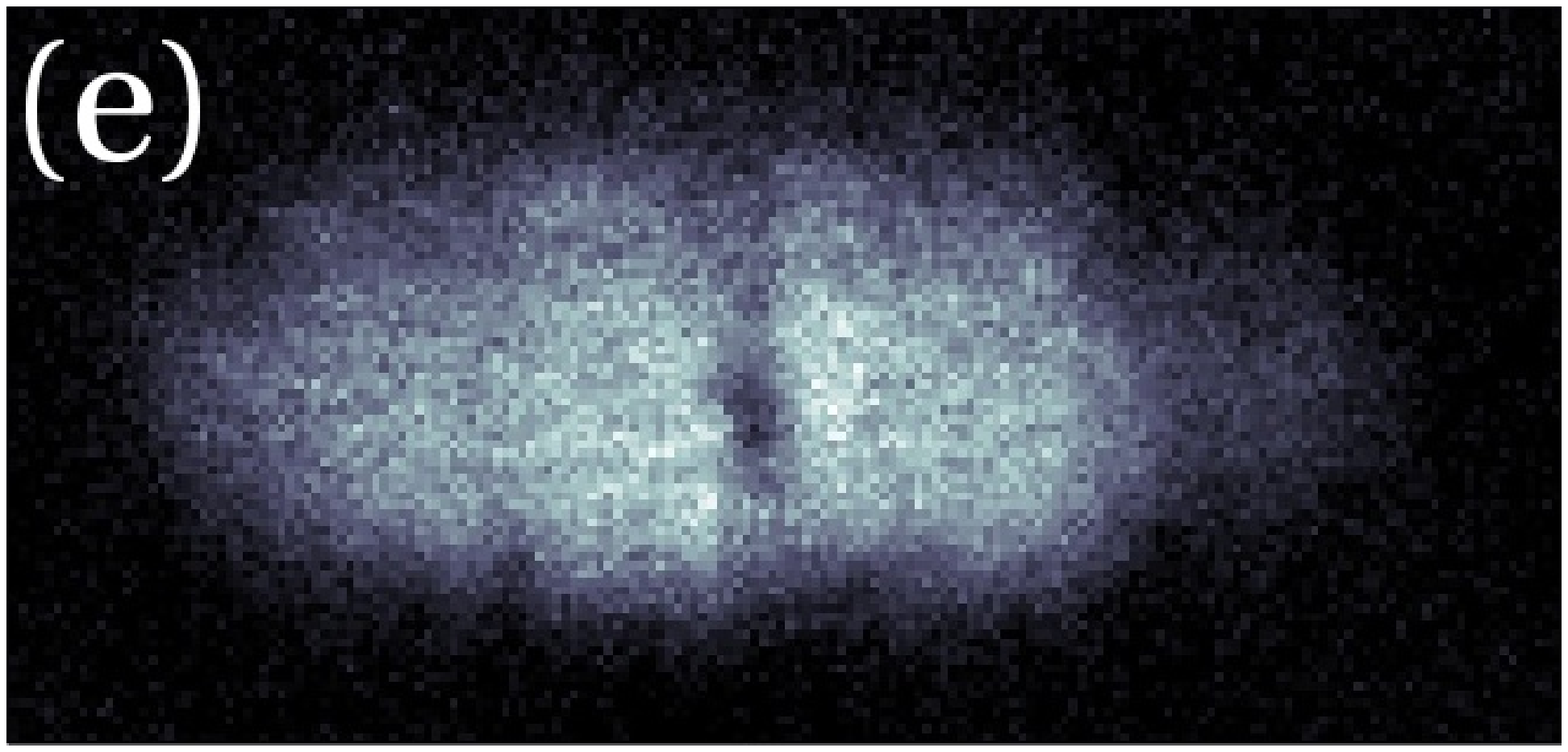}
\includegraphics[width=3.8cm]{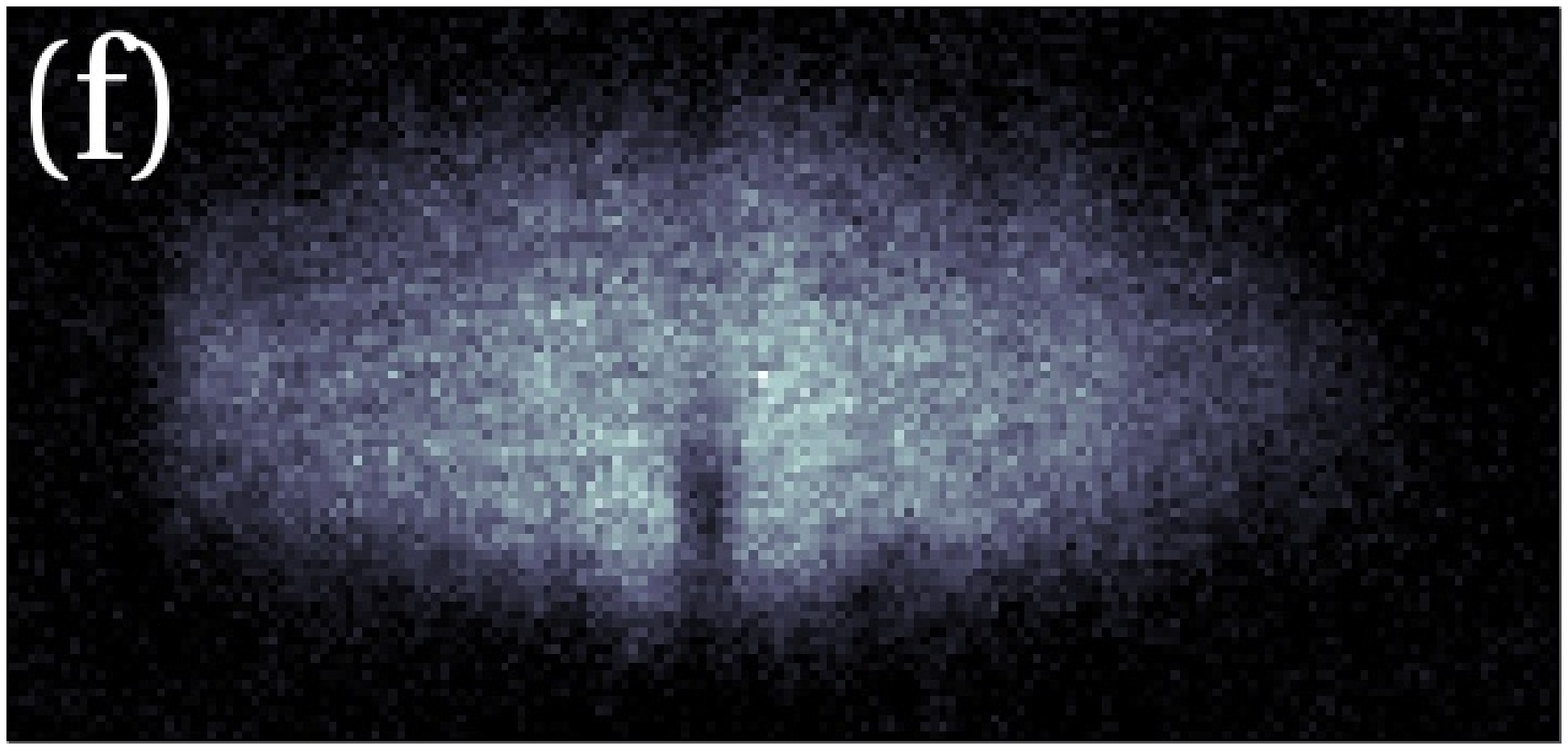}
\caption{
Samples of individual snapshots that make up the dynamics
depicted in Fig.~\ref{fig:exp2}.
The different panels correspond to
times $t=(0,12,57,119,136,155)$\,ms.
(a) Initial state containing two dark soliton stripes.
(b) The dark soliton stripes decay into vortex rings.
(c) The solitons reach the opposite edge of the
cloud and bounce back.
(d)--(f) After the strong interaction a single, stable,
solitonic vortex is left in the system.
}
\label{fig:exp3}
\end{figure}
%%%%%%%%%%%%%%%%%%%%%%%%

After closer inspection of the individual snapshots, before
integration about the transverse direction, it is evident
that the dynamics is not only governed by dark soliton stripes. This
is due to the fact that the dark soliton stripes tend to bend and
decay into vortex rings \cite{Feder:00} and also periodically
oscillate between dark soliton stripes and vortex rings \cite{jeffs}.
This behavior can be seen in the experimental snapshots
depicted in Fig.~\ref{fig:exp3}.
In particular, the first row of snapshots corresponds
to two initial dark soliton stripes that start bending [see dark
lines inside the ellipses in panel (a)]
and decay into vortex rings [panel (b)]. As the vortex rings approach
the opposite edge of the trap they recombine into a
dark soliton stripe due to the strong confinement of the trap's edge
[see dark soliton stripe very close to the right edge inside the
right-most ellipse in panel (c)].
After the ``merger'' between the two solitons a single
solitonic vortex remains in the cloud performing oscillations back and
forth along the longitudinal direction [see panels
(d)--(f)].

%%%%%%%%%%%%%%%%%%%%%%%%%%%%%%%%%%%%%%%%%%%%%%%%%%%%%%%%
\section{Model and Numerical Observations}
\label{SEC:theo}
%%%%%%%%%%%%%%%%%%%%%%%%%%%%%%%%%%%%%%%%%%%%%%%%%%%%%%%%

In an attempt to better understand the dynamics seen in the
experiments we model the BEC cloud using realistic
parameter values and initial conditions.
For this we use the 3D Gross-Pitaevskii (GP) equation that has been
shown to accurately describe the mean-field dynamics of
the BEC for low enough temperatures and large enough particle
numbers, as is the case for our setting. The GP equation, in three
dimensions, for the wavefunction $\psi(x,y,z,t)$ takes the form:
\begin{equation}
i \hbar \frac{\partial\psi}{\partial t} =
\left[ - \frac{\hbar^{2}}{2m} \nabla^{2} + V(r)
+ g_{3D} |\psi|^{2} \right] \psi,
\label{veq1}
\end{equation}
where $\nabla^2$ is the three-dimensional Laplacian,
while the trapping potential is given by
$V(r)=\frac{1}{2} m (\omega_x^2 x^2 + \omega_y^2 y^2 + \omega_z^2 z^2)$,
$m$ is the atomic mass and $\omega_i$ are the trapping frequencies
(strengths) along the different directions.
The effective nonlinearity strength is given by
$g_{3D} = {4 \pi \hbar^2 a}/{m}$, with $a$
being  the $s$-wave scattering length.
%
%It is straightforward
%to render the model of Eq.~(\ref{veq1}) dimensionless by measuring
%length, time and energy respectively in units of
%$a_y$, $\omega_y^{-1}$ and
%$\hbar\omega_y$ ($y$ is the direction of the strongest confinement
%and the corresponding harmonic oscillator length is given by
%$a_z = \sqrt{\hbar/m \omega_z}$).
%
In what follows, we will consider different initial conditions
involving initializing first a single dark soliton stripe and subsequently a few prototypical
cases corresponding to two initial dark soliton stripes.
For both of these situations we show, as a central result,
that the quasi-1D evolution will prove to be
somewhat misleading when observed as such. In all cases, it is
the true 3D dynamics (and multiple cross-sections of the BEC) that
will enable a fundamental understanding of the observations, even
when they appear entirely counter-intuitive, at first sight, as is the
case with ``plastic'' collisions.

%%%%%%%%%%%%%%%%%%%%%%%%%%%%%%%%
\begin{figure}[htbp]
\includegraphics[width=6.0cm]{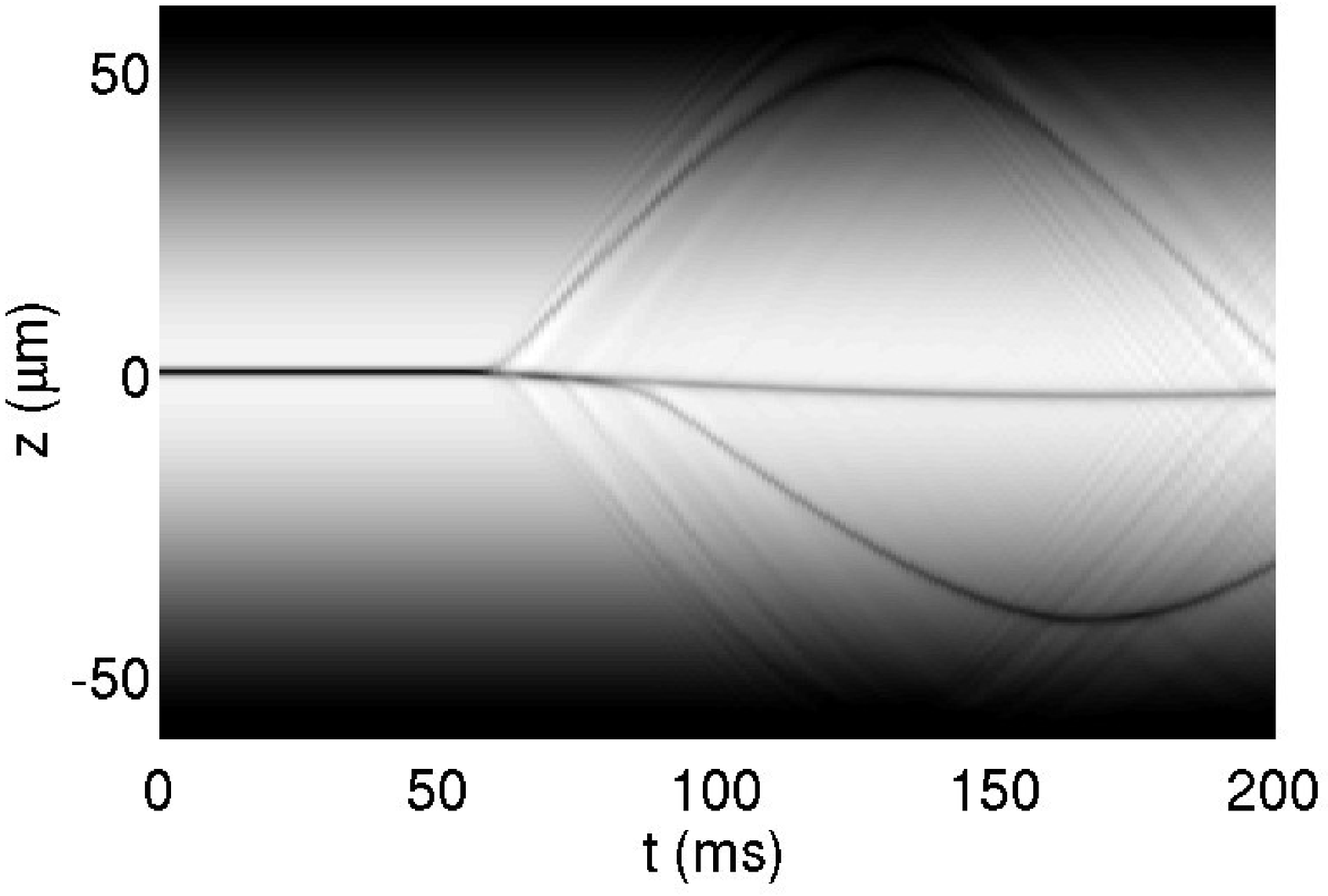}
\quad
\includegraphics[width=6.0cm]{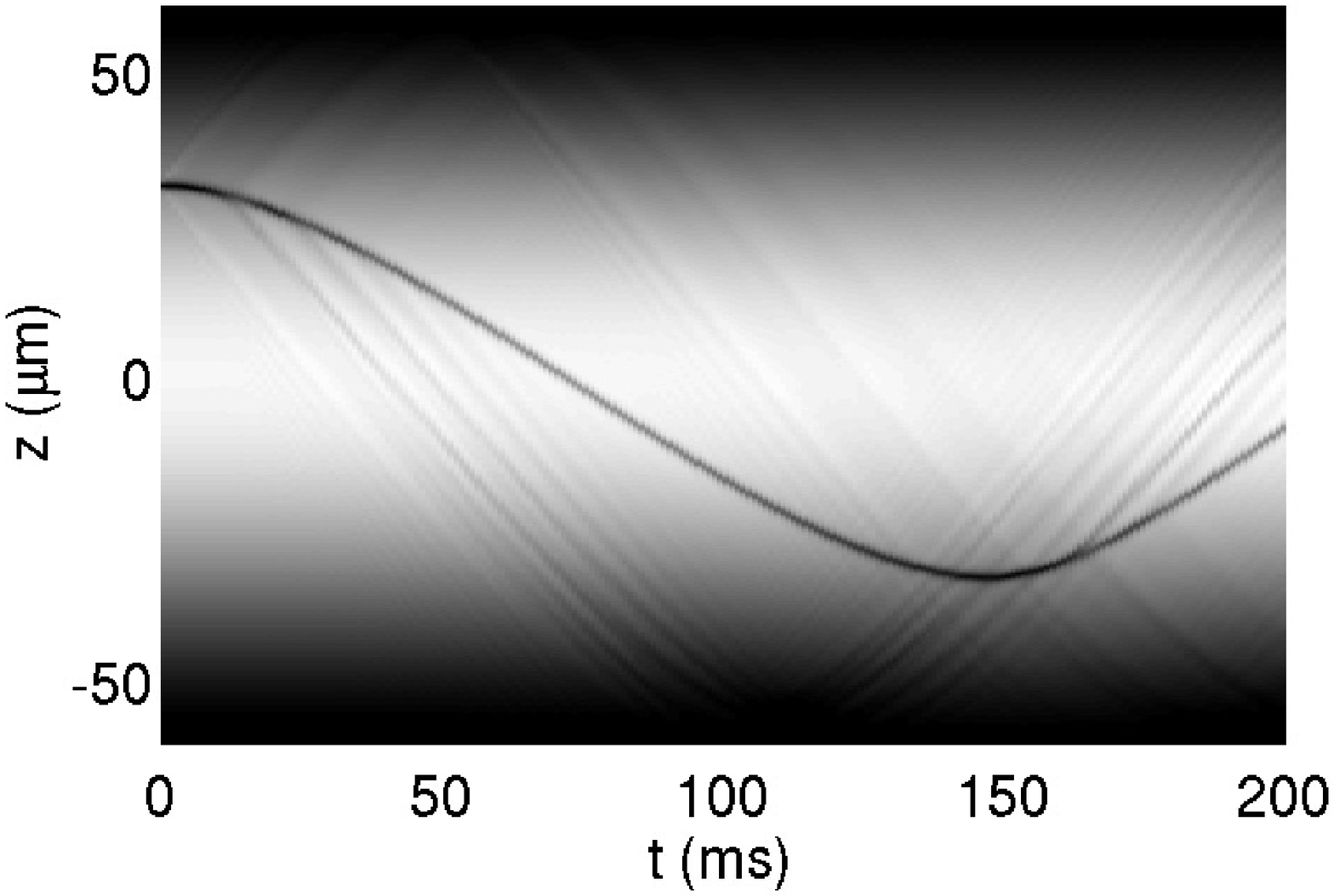}
\caption{
Left panel: Evolution of an initially stationary dark soliton stripe at the center of
the trap. The initial stationary dark soliton stripe solution is obtained by a fixed point
iterative technique (Newton method) and it is initially perturbed with a
small amount of random white noise.
The dark soliton stripe seems to split into three solitons due to its (snaking)
instability.
%
%For the respective movies of these evolutions go to the {\tt webpage}
%indicated in the abstract. There, it can be seen that the splitting
%is indeed to 3 solitonic vortices (rather than to 3 dark soliton stripes).
%
Right panel: Evolution of dark soliton stripes seeded off center.
The initial positions of the soliton is 1/2 $R_{\rm TF}$
(where $R_{\rm TF}$ is the Thomas-Fermi radius [half width of] of the cloud)
This solitary wave seems to perform stable oscillations along the
longitudinal (weak) trap direction.
All panels depict the time evolution of the density integrated
along the $x$ and $y$ directions in a manner akin to what
is done for the experimental results of Fig.~\ref{fig:exp2}.
}
\label{fig:DS_unstab}
\end{figure}
%%%%%%%%%%%%%%%%%%%%%%%%

The first thing to note about dark soliton stripes in this
trapping is that they are {\it not} always stable; see, e.g., the
discussion of Ref.~\cite{djf} and references therein.
This instability against transverse long-wavelength perturbations
was first identified in the context
of nonlinear optics~\cite{kuzne,kuz2,kuz3}, where it was also
experimentally observed
\cite{DSinstability1,DSinstability2,Kivshar-LutherDavies}.
While in a nearly spherical trap considered earlier~\cite{bpaprl3,djf},
a dark soliton stripe was shown to decay into vortex rings,
in our elongated trap experimental setup, a stationary dark soliton stripe
at the center of the trap destabilizes, due to
snaking instability, into solitonic vortices.
In fact, the left panel of
Fig.~\ref{fig:DS_unstab} depicts a numerical example\footnote{All the
numerics depicted herein were obtained using second order finite
differencing in space and fourth order Runge-Kutta integration in time.} of
the evolution ($0<t<200$) of this instability in a manner akin
to what we have observed in our experiment (cf.~Fig.~\ref{fig:exp2}).
This numerical example suggests that the initial, stationary,
dark soliton stripe at the center of the trap apparently breaks
into three dark soliton stripes which oscillate and interact.
This would obviously be a contradiction for an angular momentum
conserving (isotropic) system since any solitonic vortices have
to be nucleated in pairs of opposite charge so that total
angular momentum remains constant (equal to zero).
This also points out that in the presence of anisotropy, there
might be some internal structure of the created nonlinear waves that the
integrated density picture is failing to capture (see below).
To elucidate the true dynamics of the system it is necessary to analyze
the full 3D extent of the condensate. To that effect we will also
depict 3D isocontours of density and vorticity of the condensate.
The density corresponds to $|\psi|^2$ while the vorticity corresponds
to the curl of the fluid velocity $\vec{v}$, where
the fluid velocity is defined as the gradient of the phase of the
condensate ($\psi= |\psi|\, \exp(i \theta)$, $\vec{v} = \nabla \theta$).

%%%%%%%%%%%%%%%%%%%%%%%%%%%%%%%%
\begin{figure}[htbp]
\includegraphics[height=8.0cm]{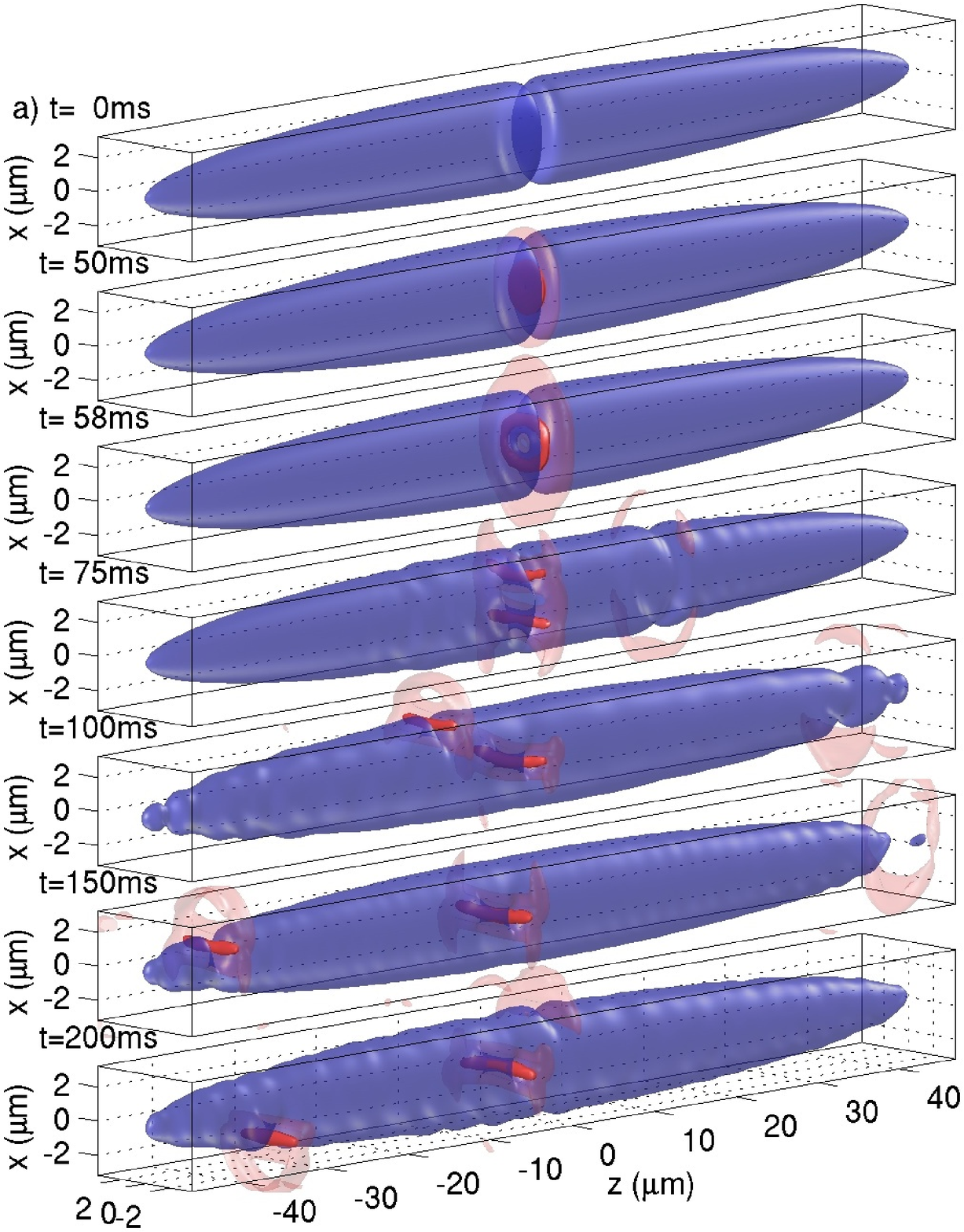}
\quad
\includegraphics[height=8.0cm]{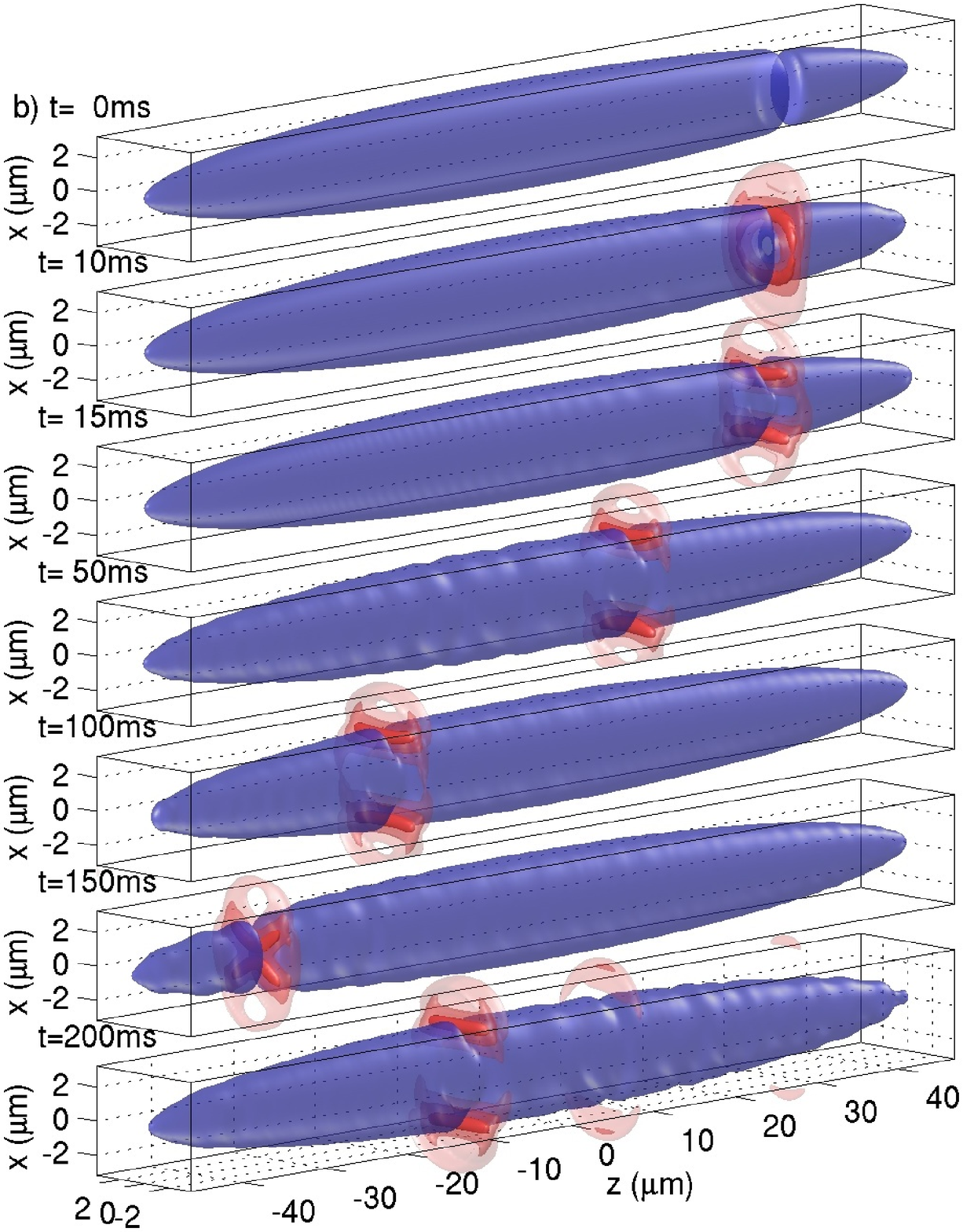}
\caption{
(Color online)
3D renderings corresponding to the numerics shown in Fig.~\ref{fig:DS_unstab}
at the times indicated.
The blue background surface corresponds to a density isocontour plot
at 40\% of the maximum density while the red surfaces correspond
to vorticity isocontours.
(a) isocontours at 50\% (light) and 85\% (solid) of maximum vorticity.
(b) isocontours at 50\% (very light), 65\% (light), and 85\% (solid)
of maximum vorticity.
}
\label{fig:1DS_3D}
\end{figure}
%%%%%%%%%%%%%%%%%%%%%%%%

Figure~\ref{fig:1DS_3D} depicts snapshots, at different times,
of density and vorticity isocontours corresponding to the examples
depicted in Fig.~\ref{fig:DS_unstab}.
As it can be observed from panel (a), it is clear that only {\em two}
solitonic vortices are created. This solitonic vortex pair corresponds to the decay of a vortex ring ($t=58$\,ms)
into a solitonic vortex pair ($t=75$\,ms) consisting of two oppositely charged solitonic vortices.
%and thus conservation of angular momentum is indeed respected.
%PGK: as we said there is no conservation to begin with, in the anisotropic case
%
Upon closer inspection, the apparent third solitonic vortex that is emitted towards
$z>0$ is actually a weak vortex ring at the periphery of the cloud. This vortex ring
is clearly visible at $t=75$\,ms, 100\,ms, and 150\,ms where it is depicted
by the light vorticity isocontours. This vortex ring is the culprit for the
``pinching'' of the BEC density at the $z$ coordinate where it lives
and thus appears as being on the same footing as the other two 
solitonic vortices in the integrated density
depicted in Fig.~\ref{fig:DS_unstab}.
For longer times (data not shown here) the ``collisions'' between
the two solitonic vortices are fairly elastic since they never get too close to
each other (one solitonic vortex remains close to the center of the trap while the
other one circles the periphery periodically),
cf.~the (asymmetric) profile of an intermediate
speed solitonic vortex in Fig.~3 of Ref.~\cite{komineas_pra}.
It is important to note that the spatial extent
in the $y$ direction is too tight to allow the coherent structure
to develop any strong excitation (or instability) in that direction
in a manner akin to the arrest of snaking instability shown
in Ref.~\cite{PanosAvoidingRedCatastrophe}.
In contrast, the confinement along the $x$-direction is weak enough
to allow for the nucleation of solitonic vortices.

\begin{figure}[htbp]
\includegraphics[width=6cm]{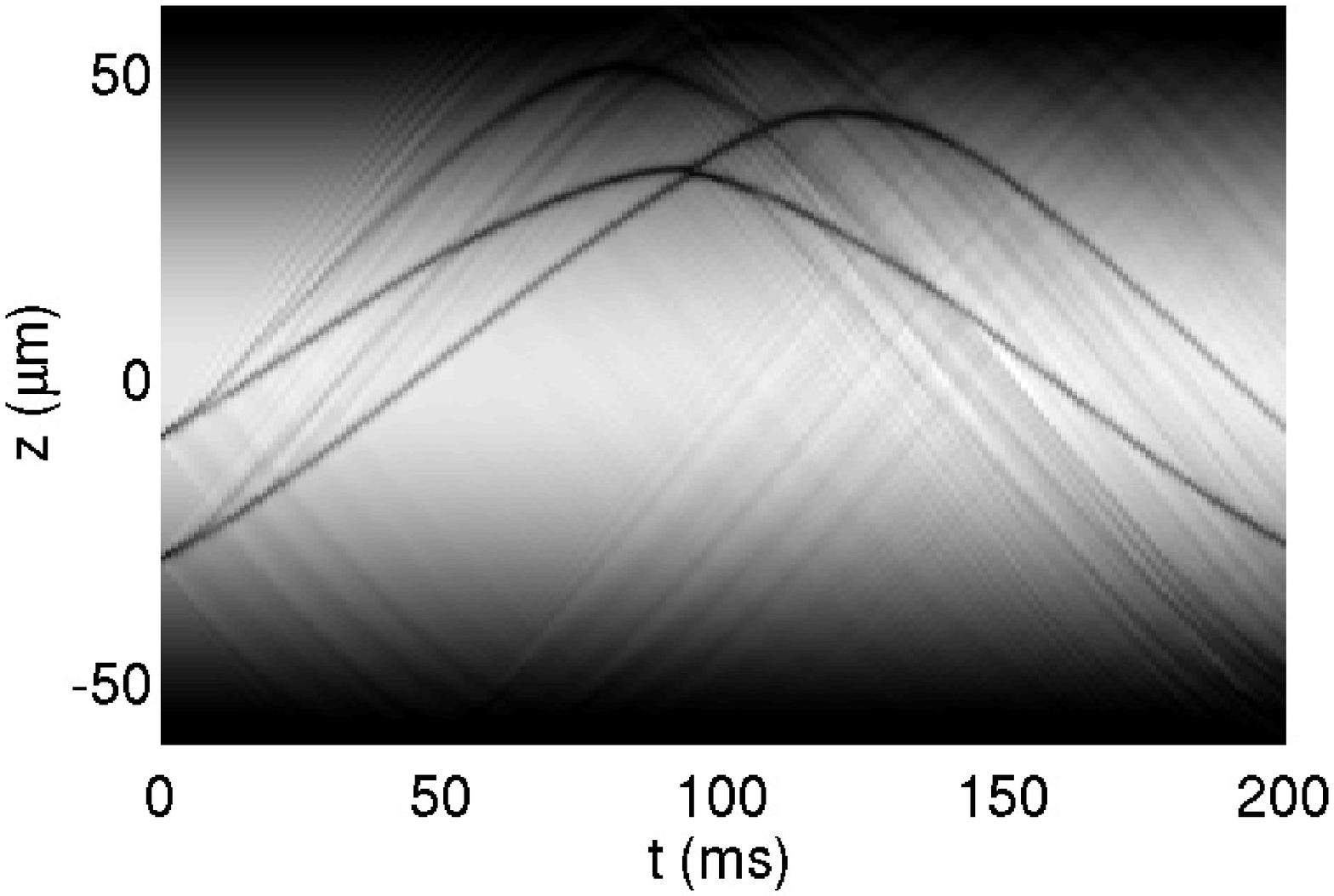}
\quad
\includegraphics[width=6cm]{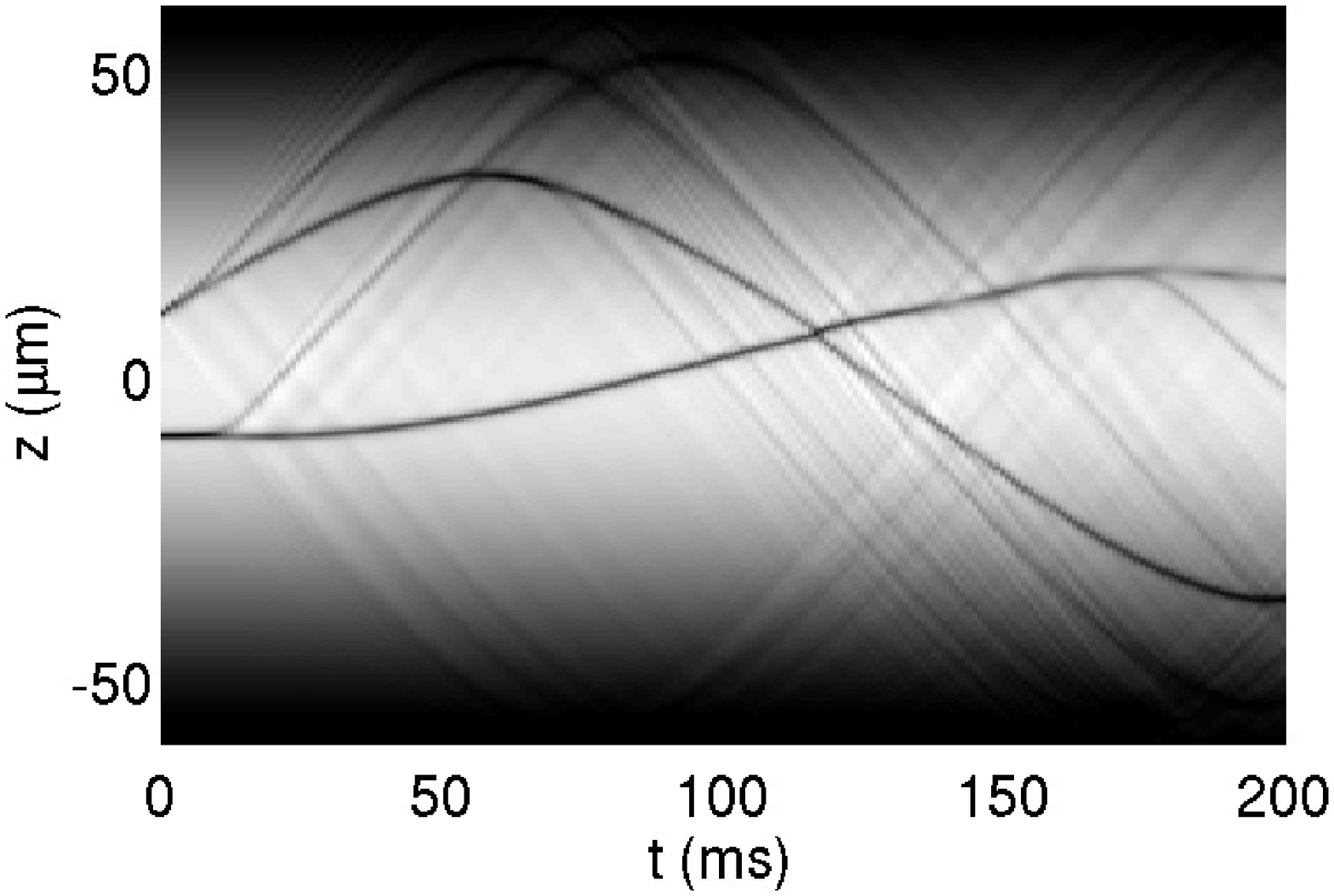}
\caption{
Evolution of two colliding dark soliton stripes.
Left: Collision when a faster dark soliton stripe
catches up with a slower one.
Right: Head on collision of a faster dark soliton stripe that bounces back from
condensate edge and collides with a slower moving dark soliton stripe.
The dark soliton stripes are indeed solitonic vortex pairs
(see Fig.~\ref{fig:2DS_3D}).
All panels depict the time evolution of the density integrated
about the $x$ and $y$ directions.
}
\label{fig:DS_collisions}
\end{figure}

We now consider the case of moving dark soliton stripes.
When the dark soliton stripes are set in motion, their instability
is reduced \cite{pelkiv}. We can see this effect in the
right panel of Fig.~\ref{fig:DS_unstab} where a dark soliton stripe
with zero initial velocity is placed a certain distance
away from the center of the trap.
As the panel shows, the dark soliton stripe placed away from the center oscillates
back and forth along the $z$ direction of the trap and thus its
instability against transverse perturbations is reduced.
This is what is also expected based on the transition
from absolute to convective instability, as was recently
discussed, e.g., in Ref.~\cite{kamchatnov}.
The specific example depicted in the right panel of
Fig.~\ref{fig:DS_unstab} corresponds to a dark soliton stripe that is initially
sufficiently far away from the center (1/2 $R_{\rm TF}$ away)
that it apparently retains its shape without splitting.
It should also be noted that this is in line with the
observation of, e.g., Fig.~4 of Ref.~\cite{komineas_pra} which suggests
that when the speeds are sufficiently large (due to the large
initial potential energy in the trap), the soliton and the solitonic vortex
merge and are indeed stabilized against the transverse modulations.
Nonetheless, the actual dynamics for the evolution of this
configuration is not fully revealed until it is depicted in 3D;
see panel (b) of Fig.~\ref{fig:1DS_3D}. As it can be observed,
the dark soliton stripe decays indeed into a vortex ring ($t=10$\,ms) but it then, in turn, appears
to quickly decay into a solitonic vortex pair ($t=15$\,ms). This solitonic vortex pair then
performs back-and-forth oscillations along the $z$ direction
of the trap.
It is crucial to note that, contrary to what was observed for the
evolution of the initially stationary dark soliton stripe placed at $z=0$
(see panel (a) of Fig.~\ref{fig:1DS_3D}), the two solitonic vortices in this
case remain {\em bound} as a pair for all times, in a way 
reminiscent of the vortex ring from which they emerged. The
binding between these two solitonic vortices is provided by a narrow (weak)
vortex ring-like vorticity structure that surrounds it. This binding
vorticity is clearly appreciated in the light vorticity
isocontours of the panel.
It is worth pointing out that the process of decay from dark soliton stripes to vortex rings and
solitonic vortices produces some surface waves in the condensate (see ripples at the
periphery of the cloud's density in all 3D renderings) that also
contain a small amount of vorticity that can be observed in some of
the vorticity isocontours corresponding to low vorticity (see for
example the light surfaces around $z=0\,\mu$m and $z=20\,\mu$m in
panel (b) of Fig.~\ref{fig:1DS_3D} for $t=200$\,ms).
These surface waves are also
visible (additional faint lines) in the evolution of the
integrated density in Fig.~\ref{fig:DS_unstab} (and also
in Figs.~\ref{fig:DS_collisions}, \ref{fig:DS_ejection} and \ref{fig:loops}).

%%%%%%%%%%%%%%%%%%%%%%%%%%%%%%%%
%%%%%%%%%%%%%%%%%%%%%%%%

%It should be noted that the oscillating dark soliton stripes depicted in the
%bottom panels of Fig.~\ref{fig:DS_unstab} are not actually
%dark soliton stripes during the whole evolution. In fact, after closer inspection
%(figures not shown here, but movies corresponding to Figs.~\ref{fig:DS_collisio%ns}
%and \ref{fig:DS_ejection} do clearly show this), the dark soliton stripes are
%actually vortex lines
%that, as they get close to the edge of the condensate,
%recombine into dark soliton stripes because of the reduce spaced in which they
%are embedded.
%Further examination of this effect is necessary to understand better
%the very nature of these moving dark soliton stripes/vortex lines.

\begin{figure}[htbp]
\includegraphics[height=8.8cm]{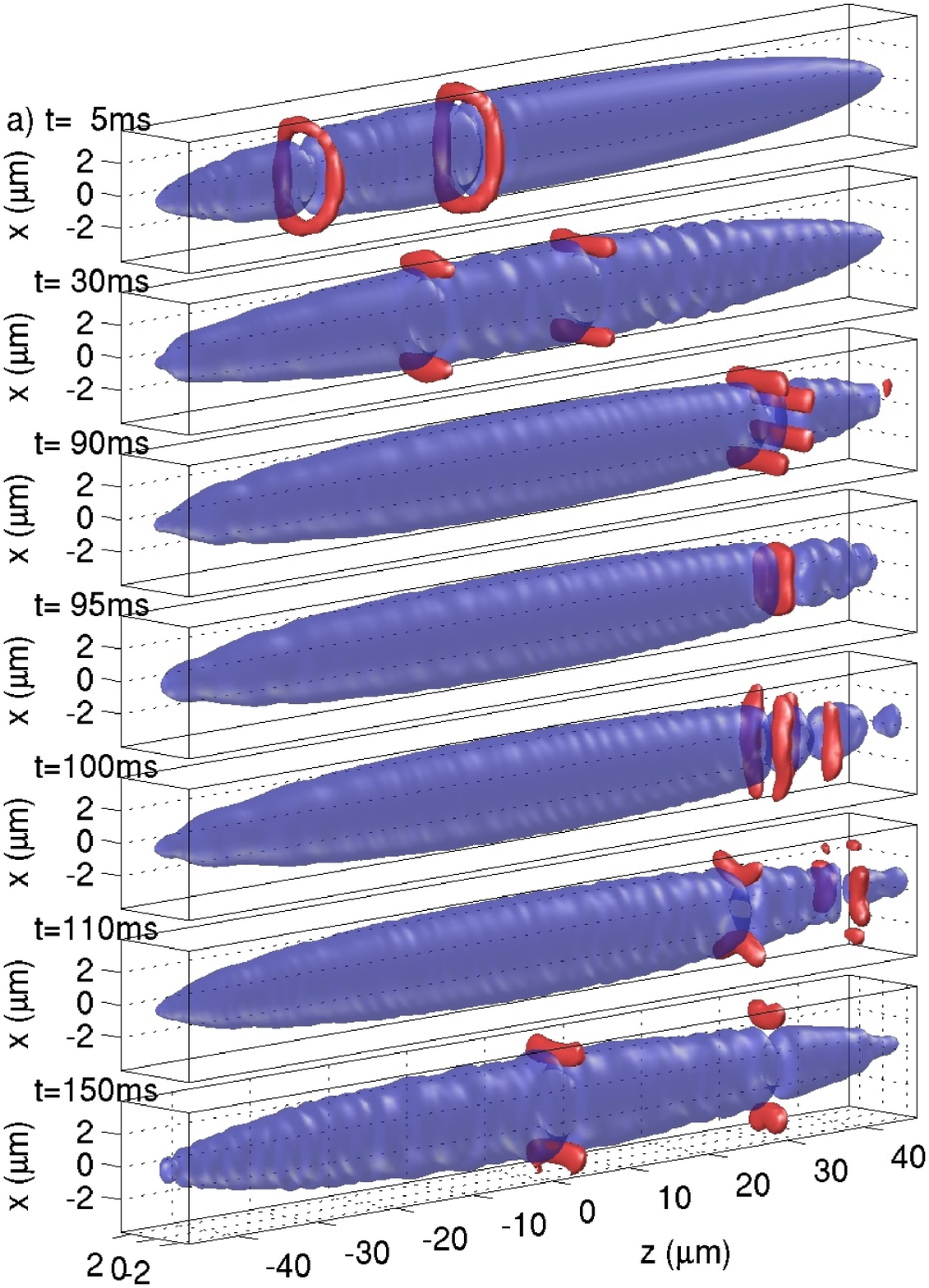}
\quad
\includegraphics[height=8.8cm]{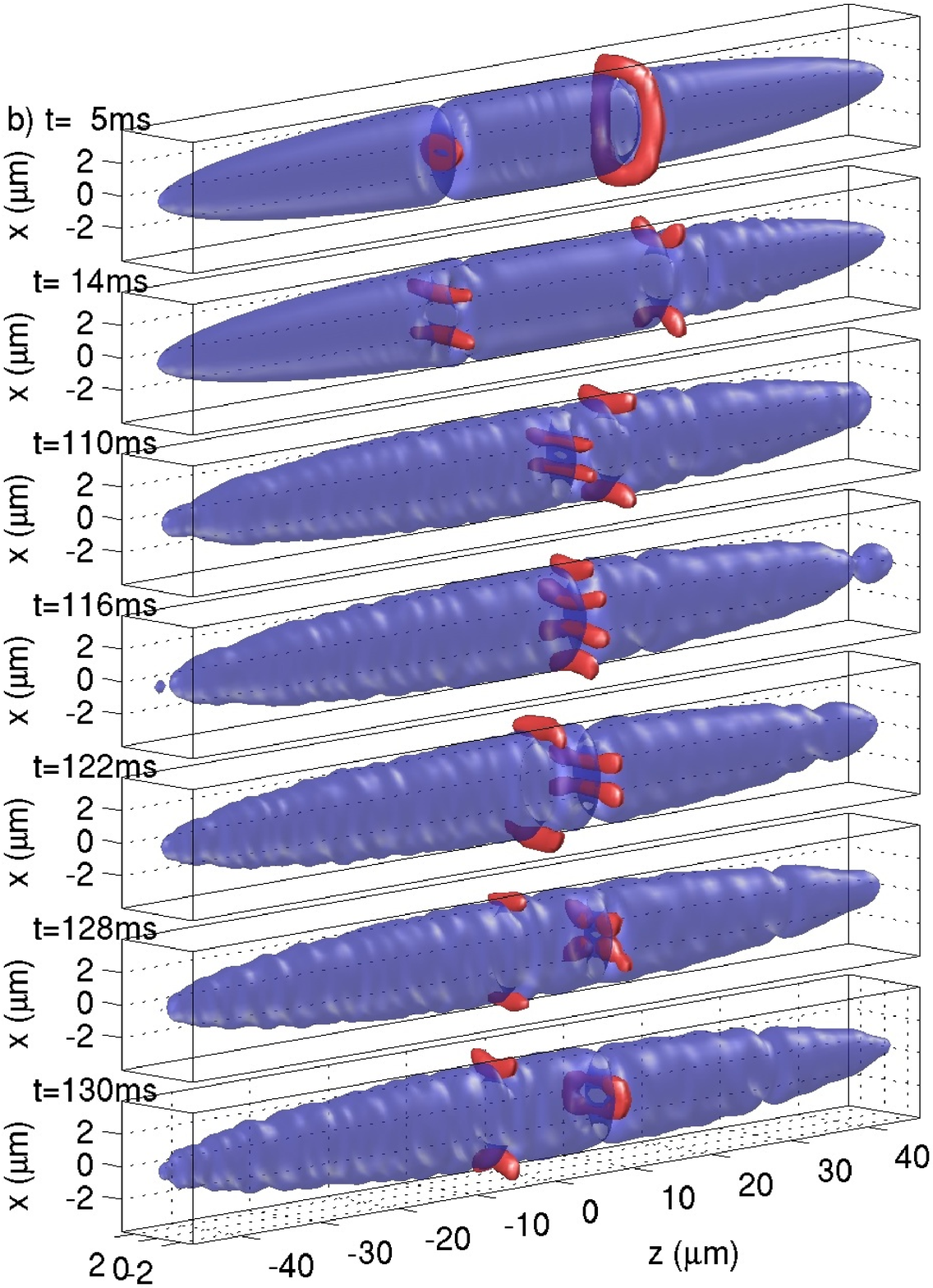}
\caption{
(Color online)
3D renderings corresponding to the numerics shown in Fig.~\ref{fig:DS_collisions}
with density and vorticity isocontours at 40\% and 85\%, respectively.
}
\label{fig:2DS_3D}
\end{figure}

\begin{figure}[htbp]
\includegraphics[width=8.0cm]{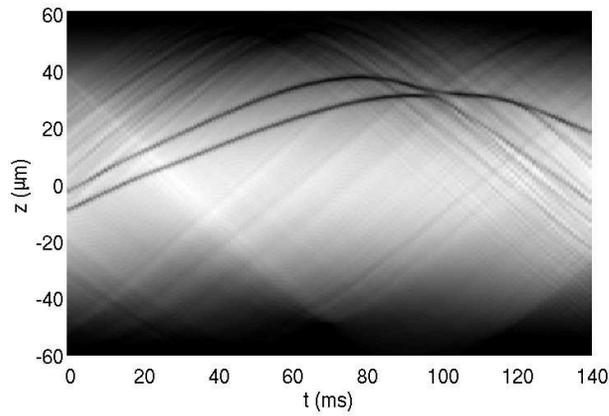}
\caption{
Numerical evolution of two chasing dark soliton stripes modeling 
conditions proximal to the experimental results shown in Fig.~\ref{fig:exp2}.
As in Fig.~\ref{fig:exp2}, we show here the time evolution of the density
integrated about the $x$ and $y$ directions.
}
\label{fig:DS_ejection}
\end{figure}

Let us now focus our attention on the collision of the localized
structures we have been describing.
In Fig.~\ref{fig:DS_collisions} we depict two examples of ``dark soliton stripe'' collisions.
The left panel shows a collision when a faster moving dark soliton stripe catches up
with a slower moving one. The right panel shows a faster dark soliton stripe that bounces
back from the edge of the condensate cloud and collides head-on
with a slower moving dark soliton stripe. In these two examples the collisions
of the dark soliton stripes seem to follow the known dynamics of interacting
and colliding dark solitons. In that line, observing such collisions seems
to suggest dynamics in close correspondence with the well-known integrable
dynamics of dark-soliton collisions~\cite{djf}. However, even in this
innocent-looking case, the dynamics is considerably more elaborate
in the present experimental setting.
More specifically, after close inspection, see corresponding
3D renderings depicted in Fig.~\ref{fig:2DS_3D}, 
the dark soliton stripes are, in reality, identified
as a pair of vortex rings ($t=5$\,ms) that subsequently appear as  two pairs of solitonic vortices that
oscillate inside the trap.
In these two cases the solitonic vortex pairs interact {\it weakly} during collision.
I.e., the 3D renderings of the condensates suggest that during the
``collision'' event, the internal separation between vortex lines
is different for the two solitonic vortices;
as a result, the interaction during collision is relatively weak,
despite the appearance of such an event in the integrated $(z,t)$-plots
of Fig.~\ref{fig:DS_collisions}.
This type of relatively weak interaction, corresponding to
seemingly elastic collisions of dark soliton stripes in the integrated density
plots of Fig.~\ref{fig:DS_collisions}, is a consequence of the
relatively large difference of the respective speeds between
the two solitonic vortex pairs as we now explain.

%%%%%%%%%%%%%%%%%%%%%%%%%%%%%%%%
%%%%%%%%%%%%%%%%%%%%%%%%

%%%%%%%%%%%%%%%%%%%%%%%%%%%%%%%%
%%%%%%%%%%%%%%%%%%%%%%%%

%%%%%%%%%%%%%%%%%%%%%%%%%%%%%%%%
\begin{figure}[htbp]
\includegraphics[width=7.5cm]{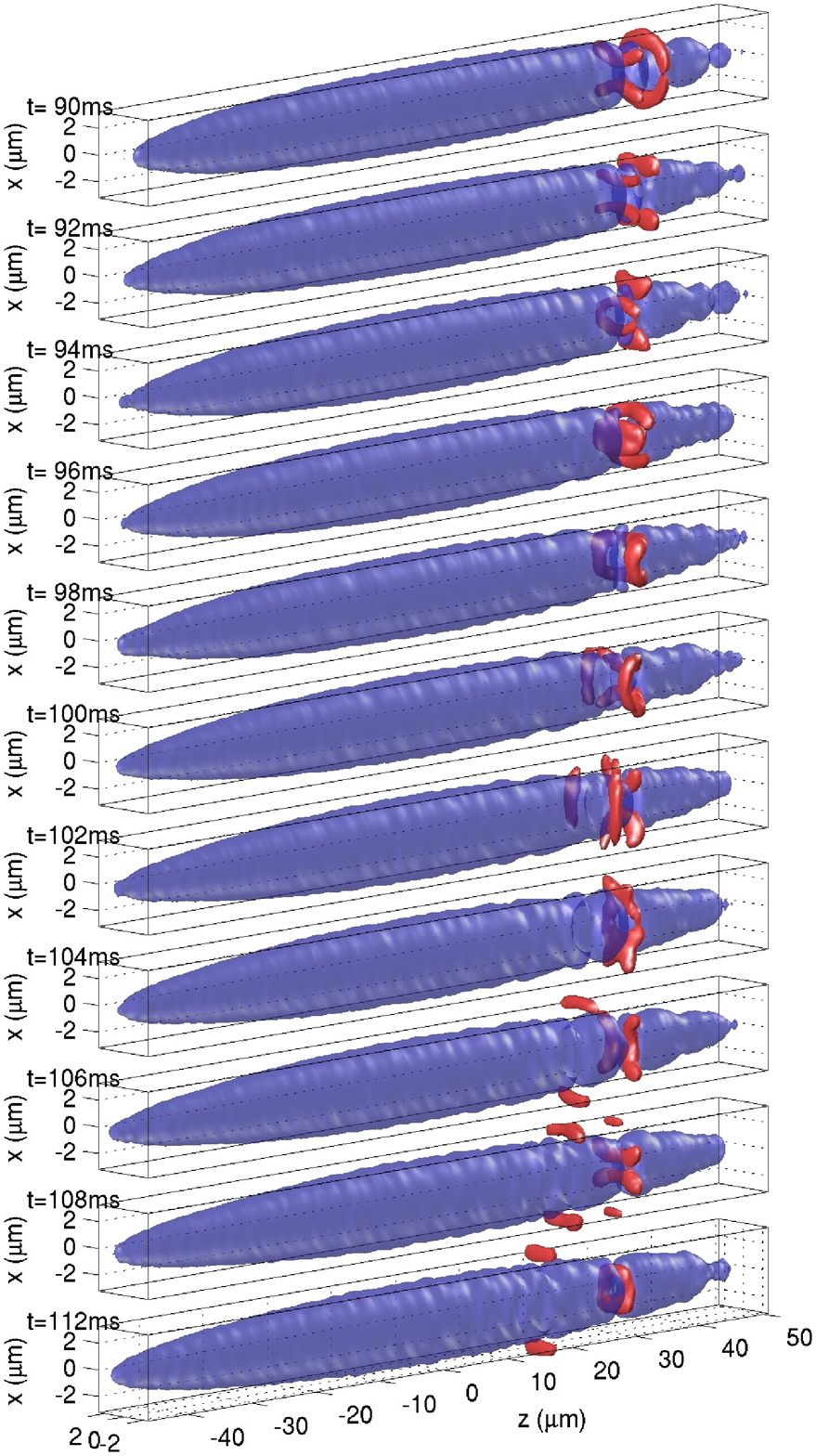}
\caption{
(Color online)
3D renderings corresponding to the numerics shown in Fig.~\ref{fig:DS_ejection}
depicting the sling shot effect suffered by one of the solitonic vortex pairs
during collision.
Density and vorticity isocontours at 40\% and 85\%, respectively.
}
\label{fig:slingshot3d}
\end{figure}
%%%%%%%%%%%%%%%%%%%%%%%%

It is well known that the speed of a vortex ring
%is, to first order,
%inversely proportional to the diameter of the ring
decreases when its diameter increases~\cite{vortex_vel}.
%This is also true for interacting 2D vortex pairs of opposite
%charge \cite{vortex_pairs}.
Therefore, our  solitonic vortex pairs (whose dynamics is strongly
reminiscent of vortex rings) traveling at
different speeds will have different inter vortex line separations
and this separation will be larger for larger difference between
their respective speeds.
This is why two solitonic vortex pairs with relatively large difference
between their speeds correspond to ones with relatively large difference
in their internal separations which can consequently pass through each
other with minimal interaction.
This effect can be observed by comparing the two panels in Fig.~\ref{fig:2DS_3D}:
panel (b) corresponds to a larger difference between the speeds of the
two solitonic vortex pairs and thus the interaction is weaker than the corresponding one of
panel (a).
Nonetheless, both of these examples can still be considered as weak
interactions as the resulting ``collisions'' seem practically
elastic.

In contrast, when the two solitonic vortex pairs have similar velocities,
the interaction is much more complex leading to unexpected behavior.
As we show now, this unexpected behavior lies at the heart of what seems
to be a merger of two dark soliton stripes in our experiment as shown
in Fig.~\ref{fig:exp2}.
Figure \ref{fig:DS_ejection} is our numerical attempt to create
conditions similar to the ones displayed by the experiment.
In this case we start with two dark soliton stripes that give rise to two
vortex rings which appear in the illustrated dynamical simulations
in the form of solitonic vortex pairs
whose velocities are relatively close.
As it can be observed from Fig.~\ref{fig:DS_ejection}, we obtain a qualitatively
similar scenario as the one observed in our experiment: the two solitonic vortex
pairs approach each other, after the outer one ``bounces back''
from the edge of the cloud, and collide ($t=100$\,ms) resulting in one
of the waves (the one closer to the center) becoming almost
imperceptible and being sling shot away at a relatively large velocity.
In Fig.~\ref{fig:slingshot3d} we depict the 3D renderings detailing
the collision between these two solitonic vortices. As it is clear from the figure,
the two solitonic vortex pairs interact very strongly because their internal separation
is almost identical. This strong interaction is responsible for one
of the solitonic vortex pairs
being sling shot away at a faster speed, leaving behind what
appears at $t=112$\,ms to clearly be a single vortex ring. Also, at the same
time, this ejected pair is relegated to the periphery of the cloud in the
$x$ and $y$ directions where the density is much weaker and so is its
imprint in the integrated density plot of Fig.~\ref{fig:DS_ejection},
resulting in a longitudinal observation  of an apparent merger of two solitonic vortex pairs.
This process is partially in line with the observation
of Ref.~\cite{kombra_prl} that intermediate collision velocities
suffer the most inelasticity, yet, to the best of our understanding
the latter work never observed numerically (or referred to experimental)
scenaria as dramatic as the one above.
This process is naturally suggestive of the fate
of the experimental evolution depicted in Fig.~\ref{fig:exp2}
where the seemingly ``plastic''
collision of two apparent dark soliton stripes (which, we essentially argue
are really solitonic vortex pairs), at around $t\approx 100$\,ms,
seems to have one of the apparent dark soliton stripes to mysteriously disappear.

%%%%%%%%%%%%%%%%%%%%%%%%%%%%%%%%
\begin{figure}[htbp]
\includegraphics[width=6cm]{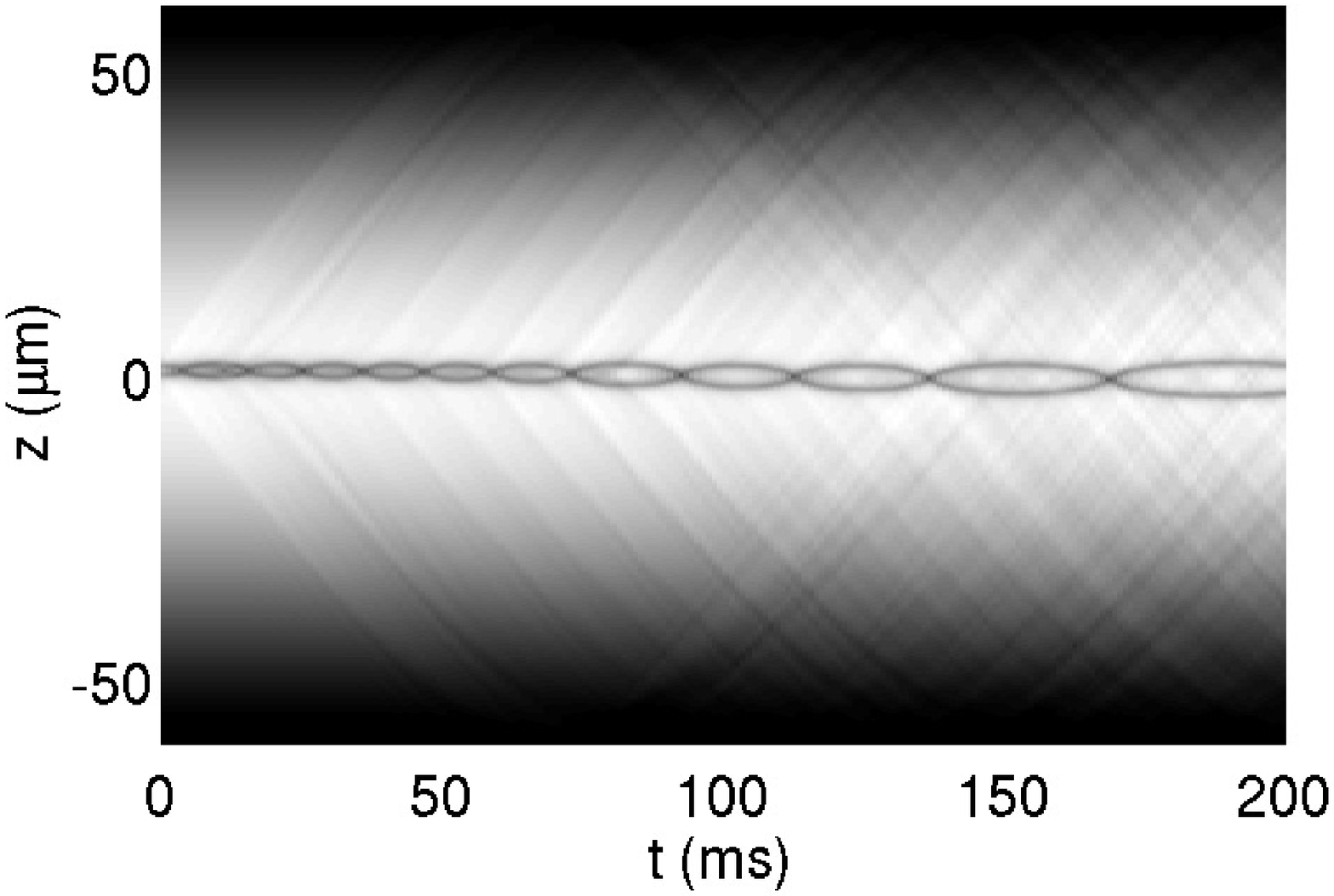}
\quad
\includegraphics[width=6cm]{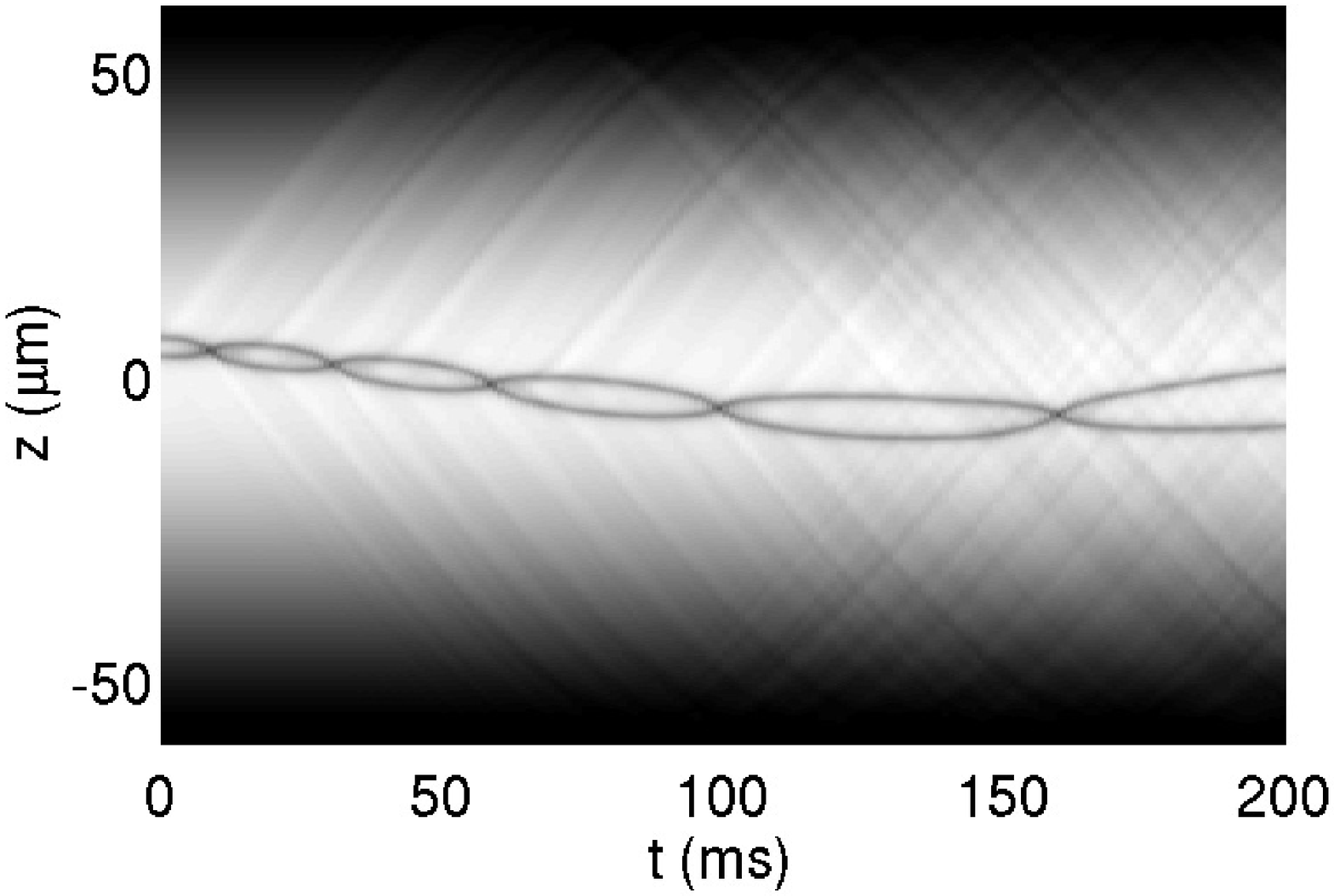}
\caption{
Evolution of two interacting seemingly dark soliton stripes (in reality, solitonic vortices).
These panels depict seemingly interacting
dark soliton stripes interacting in a highly uncharacteristic manner. These are really solitonic vortices
instead of dark soliton stripes which are initially at rest.
Left: solitonic vortices placed symmetrically about the center of the trap.
Right: same configuration but slightly displaced in the $z$ direction.
All panels depict the time evolution of the density integrated
about the $x$ and $y$ directions.
}
\label{fig:loops}
\end{figure}
%%%%%%%%%%%%%%%%%%%%%%%%

%%%%%%%%%%%%%%%%%%%%%%%%%%%%%%%%
\begin{figure}[htbp]
\includegraphics[height=7.8cm]{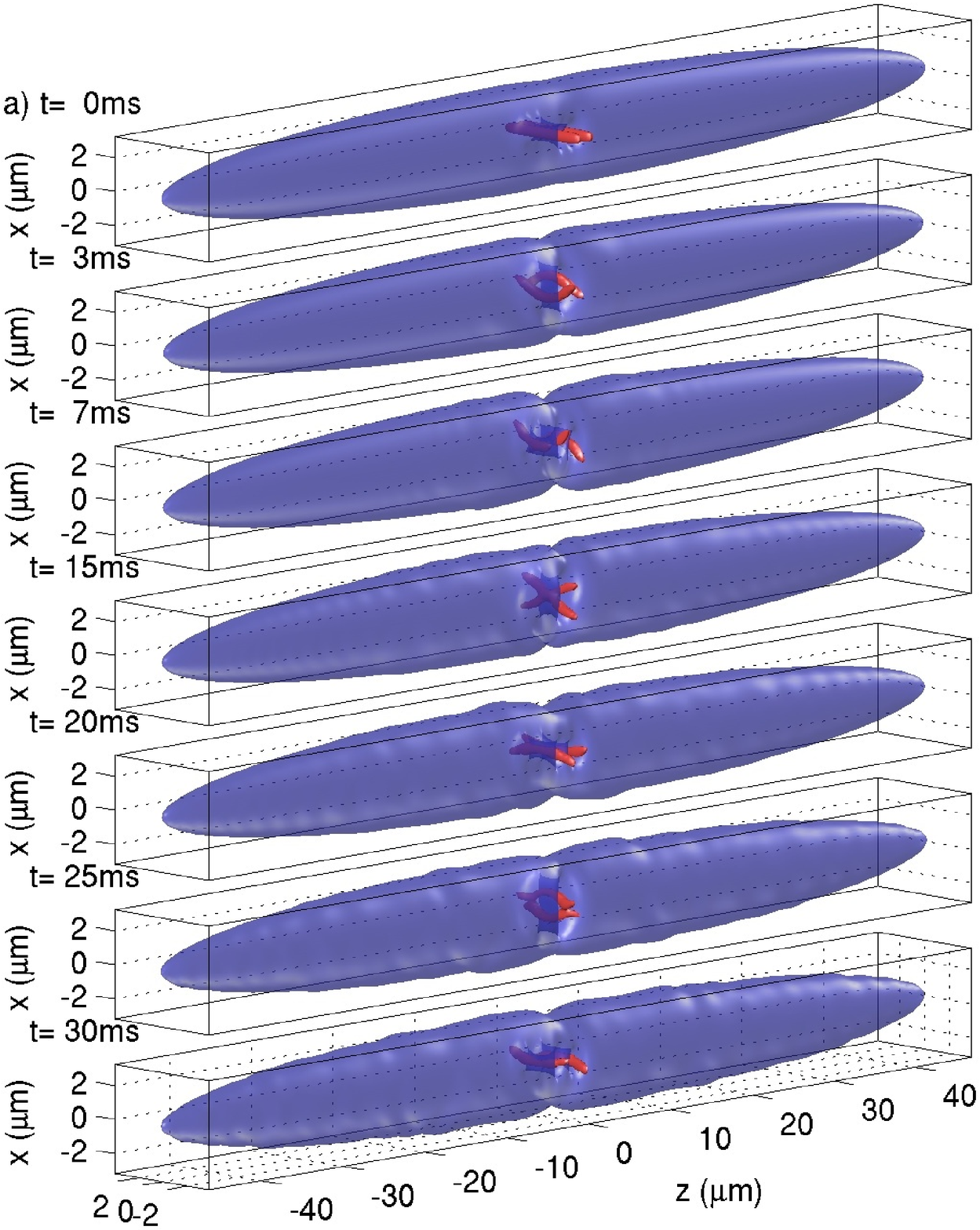}
\qquad
\includegraphics[height=7.8cm]{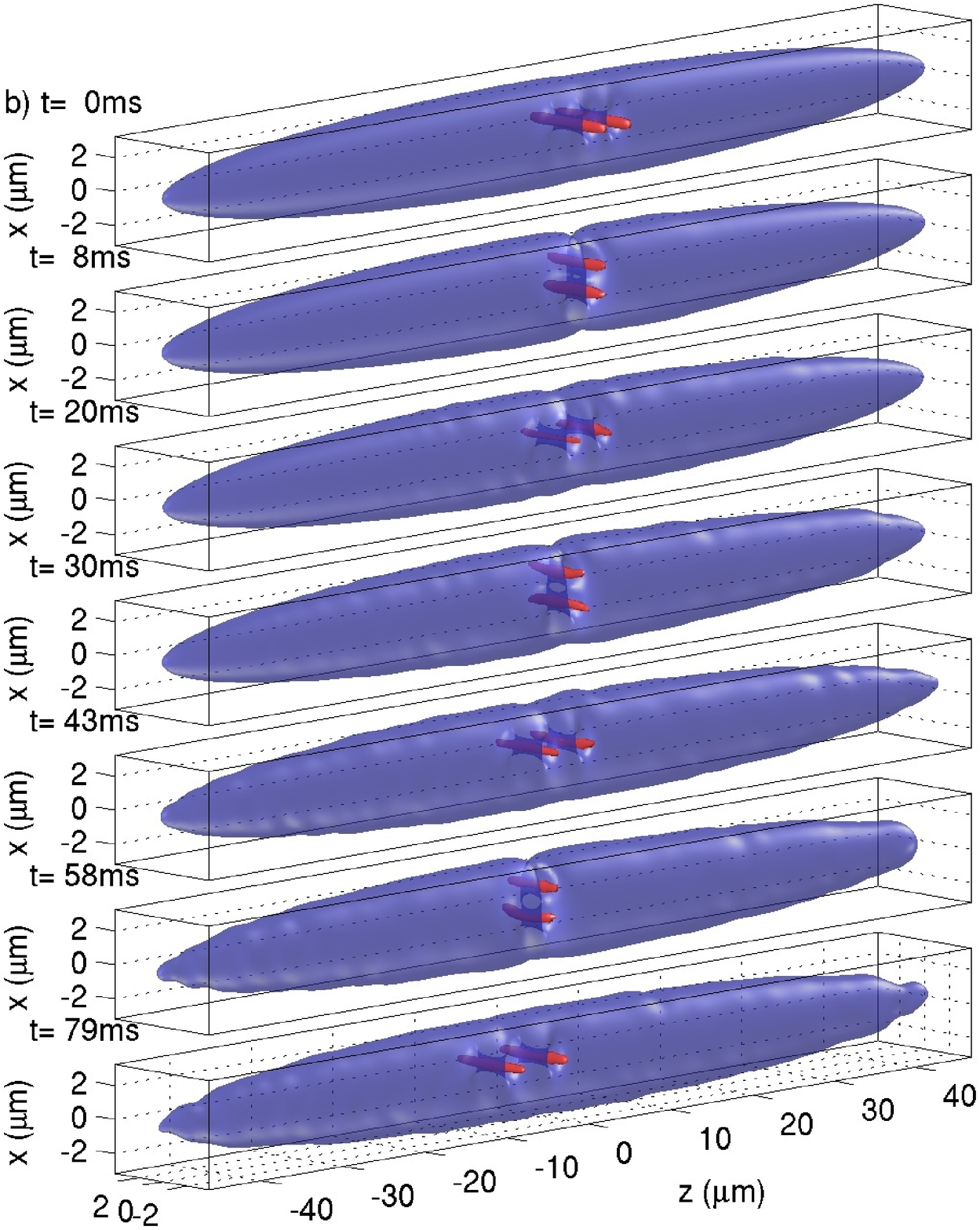}
\caption{
(Color online)
3D renderings corresponding to the numerics shown in Fig.~\ref{fig:loops}.
Density and vorticity isocontours at 40\% and 85\%, respectively.
}
\label{fig:loops_3D}
\end{figure}
%%%%%%%%%%%%%%%%%%%%%%%%

Finally, motivated by the above example,
we depict in Fig.~\ref{fig:loops} another extreme example
of an apparently inelastic collision
between dark soliton stripes. As it can be seen in the corresponding
3D renderings in Fig.~\ref{fig:loops_3D}, these
are not true dark soliton stripes but rather interacting vortex rings/solitonic vortex pairs.
In that context, the interactions
instead of arising in regular intervals as in Refs.~\cite{kip,andreas},
they instead appear to arise in intervals of increasing duration,
leading to an expanding array of ``bubbles'' (each spatio-temporal
bubble amounting to a pair of collisions).
%However, as it is clear from the
%evolution of the density integrated over the $y$ direction, we are in
%the presence of two interacting solitonic vortex pairs.
Once again, resorting to the three-dimensionality of the original
problem, we clearly observe the two solitonic vortices
interacting. Then, their vortex character (as a manifestation of
anisotropic quasi-two-dimensionality), given their same charge, enables
the possibility of rotation of the vortices around each other.
However, this rotation remains ``incomplete'' due to its confined
[anisotropic in $(x,y)$] nature. As the initial
potential energy of interaction
translates itself into rotational energy and partly gets
emitted through phonon radiation, it makes the solitonic vortices gradually get
grayer and grayer, hence more mobile within the least confined
$z$-direction. The result of this gradually increased mobility is
the more and more delayed repetition of their mutual interaction, which,
in turn, is mirrored in the expanding bubbles in the $(z,t)$-plane.

%%%%%%%%%%%%%%%%%%%%%%%%%%%%%%%%%%%%%%%%%%%%%%%%%%%%%%%%
\section{Conclusions.}
\label{SEC:conc}
%%%%%%%%%%%%%%%%%%%%%%%%%%%%%%%%%%%%%%%%%%%%%%%%%%%%%%%%
%

In this paper, by comparing experimental and numerical
results in anisotropically trapped Bose-Einstein condensates,
we show that within the ``secret lives'' of
higher dimensional dark solitons, there is more
than meets the eye in a quasi-one-dimensional inspection.
We start by considering a set of numerical investigations
with examples that exhibit either well-known dynamical
phenomena (snaking instability) or simple and innocent-looking
collisional events featuring apparently elastic interactions between
dark soliton stripes.
In each of these cases, we revealed that the true identity of
the relevant states consisted of solitonic vortices and vortex
rings. The hidden vortical nature of these higher dimensional structures
opens the possibility for unexpected dynamics and interactions.
In particular, we showcased an experimental realization that
clearly displays the manifestation of this hidden property in
a seemingly ``plastic'' collision with the (apparent) merger of
two dark soliton stripes, completely at odds with the
quasi-one-dimensional, elastic particle character of these waves.
Our numerical analogue of this experimental observation
evidences a higher-dimensional complex dynamics involving strong
interactions between two solitonic vortex pairs.
There, the apparent merger was the result of a strong interaction
between the waves whereby one of them was sling shot away.
Finally, in addition to this sling shot
event, an example of expanding oscillation bubbles with longer
times between collisions was presented and illustrated the
vortical character of the solitonic vortex interactions.
On the basis of these features, for our particular setting
of trapping strengths, the projection in the $x$ direction seems to
unveil the dynamics of dark soliton stripes since the $y$ direction is strong enough
to arrest most dynamics and any instabilities across its length. Yet, the $x$
direction is weak enough to allow for the nucleation of solitonic
vortices (vortex lines) aligned in the $y$ direction, as well as
for the illustration of vortex ring type features.
It is these extra structures, hidden in the $x$ direction, that induce
the highly atypical behavior of the apparent dark soliton stripes, as observed in the
$z$ direction.

We believe that the present investigation illustrates the
substantial value of potential further examination of interaction of
quasi-one-dimensional structures in higher dimensional and especially
in anisotropic
geometries. The context of interaction of bright solitons~\cite{expb3}
and even bright vortices in such a non-one-dimensional
context that enables radial excitations is certainly worthwhile
of further study. At the
same time the study of solitonic vortices and vortex
rings and the potential scenaria of their interaction~\cite{ginsberg}
(as well as those of the interactions between different states
or with more complex U- or S-shaped structures)
in experimentally relevant settings certainly
merits further examination and quantification, along lines
similar to what has recently been done for dark
solitons~\cite{andreas,kip} and also vortices~\cite{dsh_pra,dsh_pla}.

\section*{Acknowledgments}
We would like to thank S. Stellmer, P. Soltan-Panahi,
and S. D\"orscher for experimental support.
P.G.K. gratefully acknowledges support from the National Science
Foundation under grants DMS-0806762 and CMMI-1000337, as well as
from the Alexander von Humboldt Foundation, the Alexander S. Onassis
Public Benefit Foundation and the Binational Science Foundation.
R.C.G.~gratefully acknowledges support from the National Science Foundation
under grant DMS-0806762.
P.S. gratefully acknowledges financial support
by the Deutsche Forschungsgemeinschaft under grant Schm885/26-1.

\end{document}